\documentclass[twocolumn]{aastex63}

\usepackage{xcolor}
\usepackage{graphicx}
\usepackage{txfonts}
\usepackage{url} 
\usepackage{multirow}
\usepackage{times,epsfig,amssymb,natbib}
\bibpunct{(}{)}{;}{a}{}{,}
\usepackage{floatrow}
\newfloatcommand{capbtabbox}{table}[][\FBwidth]

\submitjournal{the Astrophysical Journal}
\received{2020 June 07}
\accepted{2020 November 30}
\shorttitle{Analysis of unconcentrated pn background}
\shortauthors{Marelli et al.}

\begin{document}

\title{Analysis of the unconcentrated background of the EPIC-pn camera on board XMM-Newton}
\correspondingauthor{Martino Marelli}
\email{martino.marelli@inaf.it}

\author[0000-0002-8017-0338]{Martino Marelli}
\affiliation{INAF-Istituto di Astrofisica Spaziale e Fisica Cosmica Milano, via A. Corti 12, I-20133 Milano, Italy}

\author[0000-0002-2483-278X]{Silvano Molendi}
\affiliation{INAF-Istituto di Astrofisica Spaziale e Fisica Cosmica Milano, via A. Corti 12, I-20133 Milano, Italy}

\author[0000-0002-9775-732X]{Mariachiara Rossetti}
\affiliation{INAF-Istituto di Astrofisica Spaziale e Fisica Cosmica Milano, via A. Corti 12, I-20133 Milano, Italy}

\author[0000-0002-9112-0184]{Fabio Gastaldello}
\affiliation{INAF-Istituto di Astrofisica Spaziale e Fisica Cosmica Milano, via A. Corti 12, I-20133 Milano, Italy}

\author[0000-0002-3853-5110]{David Salvetti}
\affiliation{INAF-Istituto di Astrofisica Spaziale e Fisica Cosmica Milano, via A. Corti 12, I-20133 Milano, Italy}

\author[0000-0001-6739-687X]{Andrea De Luca}
\affiliation{INAF-Istituto di Astrofisica Spaziale e Fisica Cosmica Milano, via A. Corti 12, I-20133 Milano, Italy}
\affiliation{INFN-Istituto Nazionale di Fisica Nucleare, Sezione di Pavia, Via A. Bassi 6, 2700 Pavia, Italy}

\author[0000-0001-7703-9040]{Iacopo Bartalucci}
\affiliation{INAF-Istituto di Astrofisica Spaziale e Fisica Cosmica Milano, via A. Corti 12, I-20133 Milano, Italy}

\author[0000-0002-3758-9272]{Patrick K\"uhl}
\affiliation{Christian-Albrechts-Universit\"at zu Kiel, Leibnizstr. 11, 24118 Kiel, Germany}

\author[0000-0003-4849-5092]{Paolo Esposito}
\affiliation{INAF-Istituto di Astrofisica Spaziale e Fisica Cosmica Milano, via A. Corti 12, I-20133 Milano, Italy}
\affiliation{Scuola Universitaria Superiore IUSS Pavia, piazza della Vittoria 15, 27100 Pavia, Italy}

\author[0000-0003-0879-7328]{Simona Ghizzardi}
\affiliation{INAF-Istituto di Astrofisica Spaziale e Fisica Cosmica Milano, via A. Corti 12, I-20133 Milano, Italy}

\author[0000-0002-6038-1090]{Andrea Tiengo}
\affiliation{INAF-Istituto di Astrofisica Spaziale e Fisica Cosmica Milano, via A. Corti 12, I-20133 Milano, Italy}
\affiliation{INFN-Istituto Nazionale di Fisica Nucleare, Sezione di Pavia, Via A. Bassi 6, 2700 Pavia, Italy}
\affiliation{Scuola Universitaria Superiore IUSS Pavia, piazza della Vittoria 15, 27100 Pavia, Italy}

\begin{abstract}

Our understanding of the background of the EPIC/pn camera onboard {\it XMM-Newton} is incomplete. This affects the study of extended sources and can influence the predictions of the expected background of future X-ray missions, such as {\it ATHENA}. Here, we provide new results based on the analysis of the largest data set ever used. We focus on the {\it unconcentrated} component of the EPIC/pn background -- supposedly related to cosmic rays interacting with detector and telescope structures. We show that the so-called out-field of view region of the pn detector is actually exposed to the sky. After carefully cleaning from the sky contamination, the unconcentrated background measured in the out-field of view region does not show significant spatial variations and its time behaviour is anti-correlated with the solar cycle. We find a very tight, linear correlation between unconcentrated backgrounds detected in the EPIC/pn and the EPIC/MOS2 cameras. This relationship permits the correct evaluation of the pn unconcentrated background of each exposure on the basis of MOS2 data, avoiding the use of the contaminated out-field of view region of pn, as done in standard techniques. We find a tight linear correlation between the pn unconcentrated background and the proton flux in the 630--970 MeV energy band, as measured by the EPHIN instrument on board SOHO. Through this relationship we quantify the contribution of cosmic ray interaction to the pn unconcentrated background. This reveals a second source which contributes to the pn unconcentrated background for a significant fraction (30\%-70\%). This agent does not depend on the solar cycle, does not vary with time and is roughly isotropic. After having ruled out several candidates, we find that hard X-ray photons of the CXB satisfy all the known properties of the constant component. Our findings provide an important observational confirmation of simulation results on ATHENA and suggest that a high energy particle monitor could  contribute decisively to the reproducibility of the background for both experiments on ATHENA.
   
\end{abstract}


\section{Introduction} \label{intro}

The {\it XMM-Newton} observatory \citep{jan01} has been providing for the last 20 years the largest collecting area of any imaging X-ray telescope flown to date.
This large collecting area is exploited by the European Photon Imaging Camera (EPIC) detector composed by a pn \citep{str01}
and two MOS CCD cameras \citep{tur01} at the foci of its three X-ray telescopes.
The EPIC capabilities to characterize low surface brightness emission from extended and
faint objects such as supernova remnants and clusters of galaxies are highly affected -- more than expected before the launch -- by a high and variable background \citep[e.g.][]{lum02,del04,rea03} that is still poorly understood. The characterization of low surface brightness emissions is also amongst the top science goals for {\it ATHENA} (Advanced Telescope for High Energy Astrophysics), the future large mission of ESA dedicated to the X-ray band to be launched in the early 2030s.
To achieve this goal, it is vital to minimize the intensity of the background that will affect {\it ATHENA} and maximize its reproducibility during the current phase of mission design, orbit evaluation and definition of the observational strategy. 
As discussed in \citet{mol17}, much can be learnt from previous X-ray mission. In this respect, {\it XMM-Newton} is particularly useful both because its EPIC pn camera bears strong similarities with the {\it ATHENA} WFI instrument and because during most of its highly elliptical orbit it is not shielded by the Earth magnetic field and thus is affected by high-energy galactic cosmic rays as {\it ATHENA} will be in its orbit around L1 or L2\footnote{L2 is the original baseline orbit for {\it ATHENA}. Based on the analysis of the available data on soft protons at L1 and L2, recently the Athena Science Study team recommends an L1 orbit (Guainazzi 2020, ASST-SOM-0014). The final decision will be taken within few months.}.

The EPIC background can be divided into three main components: the background due to noise and defects of the detectors, the component concentrated by mirrors (X-ray photons and low-energy protons), and the unconcentrated component (mainly consisting of high-energy particles and their secondaries).\\
The detector background comprises electronic noise (e.g. readout noise, dark current) and damaged pixels/columns. The first part is important at low energies (i.e. below 300 eV) while the second part is limited to specific small areas of the detector (bright pixels and columns).\\
The concentrated background is produced by two components: photons and soft protons. The photon background is due to the emission of unresolved AGN (the cosmic X-ray background, CXB), of unresolved galactic and extended sources and of the Galaxy. Their overall rate is almost time-independet and it depends only on the sky region within the field of view (FOV). Low-energy protons
(with energies smaller than a few 100 keV) are funneled towards the detectors by the X-ray mirrors and detected as events; this component is characterized by strong (up to three orders of magnitude) and rapid (down to few tens of seconds) variability.\\
The unconcentrated background is mainly due to the interaction of high-energy particles (typically cosmic rays with energies larger than some 100 MeV) with the structure surrounding the detectors and with the detectors themselves. Part of the particles (e.g. the minimum ionizing particles, MIP) are recognized and the events ascribed to them are rejected onboard the satellite, resulting in time-dependent excluded CCD columns and/or events with invalid patterns (i.e., illuminating nearby pixels spatially distributed differently than expected for valid X-ray events).

The knowledge of these components has been improving thanks to the many efforts involved in collecting suitable blank sky fields to be used as template background
by the {\it XMM-Newton} users \citep{rea03,car07}, performed by the {\it XMM-Newton} NASA Guest Observer Facility\footnote{https://heasarc.gsfc.nasa.gov/docs/xmm/xmmgof.html} leading to the
{\it XMM-Newton} Extended Source Analysis Software \citep{sno08,kun08}, to the efforts of the {\it XMM-Newton} Scientific Organizing Committee (SOC) and the contributions of various research teams.
Our team has been particularly active on this topic \citep{del04,gas07,lec08} and we have recently exploited
a large data set of MOS2 observations to advance further the knowledge of the {\it XMM-Newton} focused and unfocused background components \citep{mar17,sal17,ghi17,gas17}. These latter works were developed in the framework of the R\&D ESA project AREMBES (ATHENA Radiation Environment Models and X-Ray Background Effects Simulators) and contributed to estimating the expected background level of ATHENA detectors and to favouring the baseline orbit around L1 rather than L2.\\
All these studies were focused mainly on the MOS cameras. A key feature of 
these detectors is the presence of portions that are not exposed to the sky (out field of view, outFOV), so neither celestial X-ray photons, nor soft protons focused by the mirror are collected there. This continuous monitoring of the unconcentrated background allows to evaluate such component also in the focused area, where celestial photons and soft protons dominate.\\
The pn background has been studied in \citet{kat04,fre04,fra14}. The differences that make the pn analysis more difficult with respect to that of the MOS are the relatively small outFOV
detector area -- about a quarter than of MOS2 -- and the higher percentage of Out of Time (OoT) events (from 2.3\% to 6.3\% for the pn as opposed to 0.35\% for MOS). Such effect is due to the CCD transfer time: a minor fraction of in-field-of-view (inFOV) events is incorrectly assigned to the outFOV region as OoT events.

With these differences in mind, our team tackled the issue of studying the unconcentrated background in the pn camera making use of events falling in the outFOV, as we did for the MOS2 camera. To this aim, we took advantage of the big data sets available within the R\&D ESA project AREMBES and the FP7 European project EXTraS\footnote{http://www.extras-fp7.eu/} \citep[Exploring the X-ray Transient and variable Sky,][]{del16}.
Unfortunately, it was clear from the inspection of the first data sets that the outFOV areas of the pn are
contaminated by the focused components, both photons and soft protons. In Section \ref{cont_pre} we demonstrate the presence of this contamination using small sub-samples of the available {\it XMM-Newton} observations. Here we give a general idea of the characteristics and origin of this contamination, that will be useful throughout the paper to minimize it in our larger sample.
We present the data set we use and the event selection we adopted in Section \ref{data}, with a particular care for the minimization of the contaminants (details about the contamination effects and the minimization of such component in our event set are described in Appendix \ref{off_script} and \ref{app:cont}). In Section \ref{data_analysis}, we present our analysis to characterize the temporal behaviour of pn unconcetrated background. In Section \ref{filtermode}, we study the resulting time series, investigating whether different choices of filters and observing mode affect pn background, as well as the correlation between pn and MOS2 unconcentrated backgrounds. In section \ref{imaging}, we report on the analysis of the spatial distribution of the pn unconcentrated background within the outFOV. Finally, in Section \ref{disc}, we discuss the results of our analysis, also correlating them with the contemporaneous time series of high energy protons accumulated with EPHIN detector onboard SOHO \citep{muller95,kul16}. In Section \ref{summary}, we summarize the main results of this paper.

\section{The problem of contamination in the pn corners} \label{cont_pre}

Our study of the unconcentrated background of the pn camera is based on events falling in its outFOV area.
Since this is the first time that such a study is performed, we need to check if other background components contaminate the pn outFOV, such as concentrated photons and/or soft protons (SP). Making a first analysis on small samples, we are able to obtain a zero-order characterization of contamination(s) and to choose the best techniques to minimize it in larger samples.

We selected two different case studies: one with a large amount of SP and the other with a large amount of photons distributed over the entire inFOV.
The first case study ("proton sample") is composed by a selection of long {\it XMM-Newton} observations\footnote{ObsIDs 0210490101, 0302500101, 0305540701, 0305541101 and 0305540601} contaminated by SP, with a total exposure of 504 ks.
No bright extended sources are in the field of these observations, and the number of point-like sources is limited.
For this sample we select and use only periods of high and variable background to maximize the SP contamination.
The second case study ("photon sample") comprises the {\it XMM-Newton} observations\footnote{ObsIDs 0124710401, 0124710601, 0124710901, 0124711101 and 0124711401} pointing to the center of the Coma cluster, one of the brightest, hardest and most extended
X-ray sources in the sky, with a total exposure of 115 ks.
For this sample we select and use only periods of low and constant background so to exclude the SP contamination and magnify the photons' contamination.\\
The periods of high soft protons background are in this respect not exceptional as they represent 40\% of the {\it XMM-Newton} time available to observation as many studies reported and confirmed on a large dataset by our work \citep{sal17}. On the conrary the photon contamination in the pn outFOV by the observation of the central regions of Coma are rather exceptional.

\begin{figure*}[hbt!]
\centering
\includegraphics[width=0.75\textwidth]{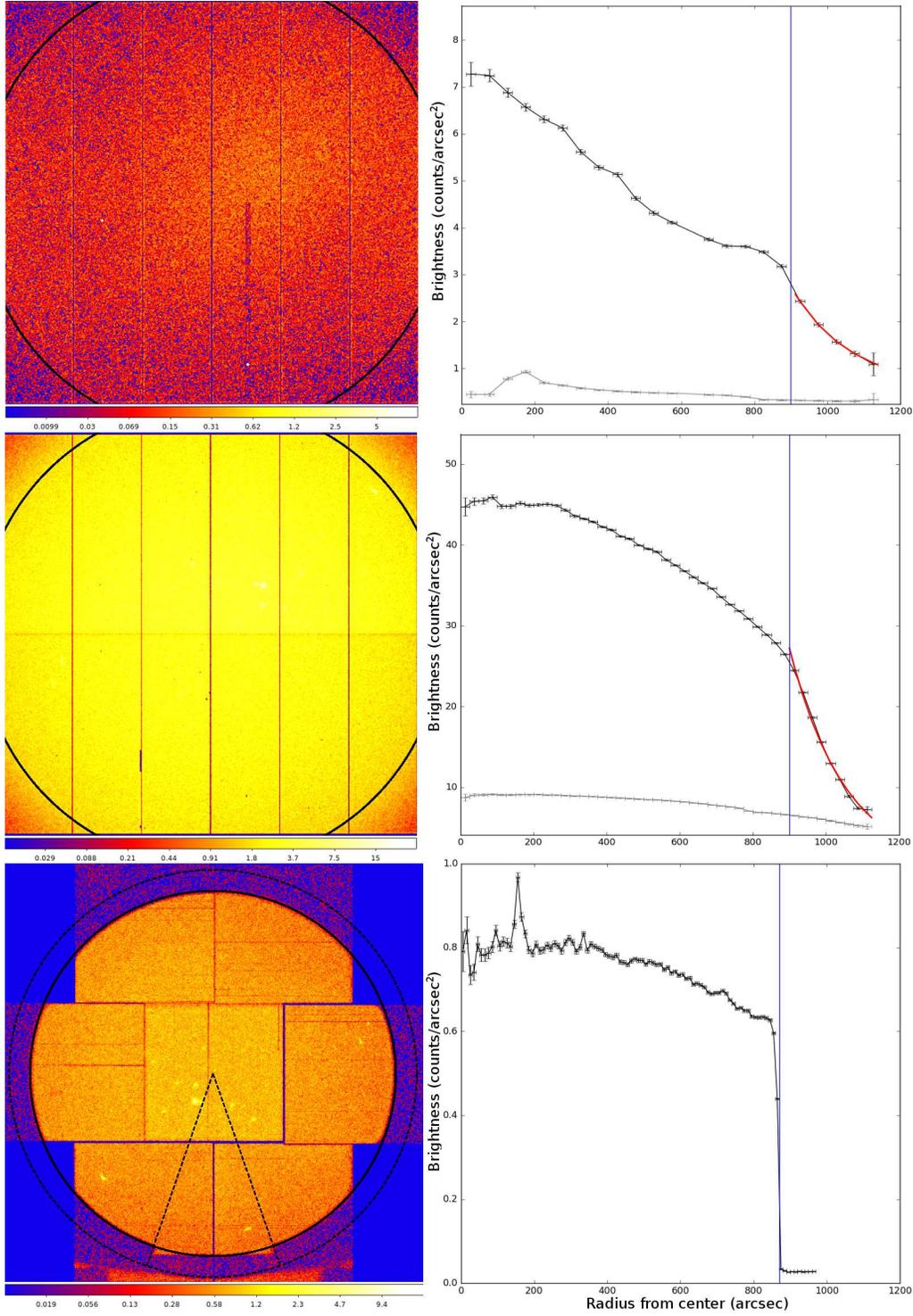}
\caption{Count map (on the left) and radial surface brightness profiles (on the right) of the two pn case studies we analyzed in Section \ref{cont_pre}: photons sample (upper panels) and
  protons sample (middle panel). We also show the count map and radial profile obtained for MOS2 camera using the protons sample (lower panel); in this case, we excluded from the radial profile analysis
  the regions outside the dashed circle and within the dashed sector due to the deviation from a circular geometry of the inFOV \citep[see also][]{mar17}.
  The border between inFOV and outFOV regions is marked by a black circle in the images and a blue vertical line in the brightness
  profiles. In the brightness profiles (on the right), we show the area-normalized count rate (black points) and, for the two pn data sets, the predicted out-of-time events (grey points). In red, we also show the best fit of the pn outFOV region using, as a first approximation, a constant plus exponential model.
}
\label{fig_preliminary}
\end{figure*}

We followed standard analysis procedure for pattern and flag selection \citep[see e.g.][]{ros16}. We evaluated the contamination by OoT events following the procedure that will be discussed in Section \ref{oot_rescale}. For each exposure we extracted the count maps in detector coordinates (1 pixel = 1 arcsec$^2$). Maps from different observations were summed and
for each map we produced a radial brightness profile centered in the geometrical center of the pn detector.\\
Formally, the inFOV area is defined as a circle with a 900" radius from the geometrical center of pn, while the remaining area is outFOV \citep{str01}. From Figure \ref{fig_preliminary}, it is apparent that the entire outFOV area is contaminated by photons (upper panels) and SP (central panels). The contamination is larger in the inner part of the outFOV and decreases following, as a first approximation, a quasi-exponential model. This is completely different with what observed in the case of the MOS cameras: we show as an example the MOS2 view of the "proton sample" in the lower panel of Figure \ref{fig_preliminary}. The sudden drop in the count-rate by more than a factor 10 at the border of the FOV is apparent.

Since for pn there is no region unaffected by concentrated background, the use of the term outFOV to describe the corners of the pn is something of a misnomer, as these regions are sensitive, albeit only partially, to SP and photons. However, faced with the choice of either adopting a new term or maintaining the term "outFOV", we opt for the latter.\\
Also, it is obvious that to extract the unconcetrated background we cannot use all the events falling in the outFOV region as we did for the MOS2 camera \citep{mar17}.
For the analysis of the larger data set, we will adopt two different filters to mitigate the contamination by photons and SP in the outFOV region (see Section \ref{data}). Also, the residual contamination will be evaluated a posteriori at the end of the analysis (Sections \ref{pnmos2}, \ref{imaging}).

\section{Data Preparation} \label{data}

In this section, we describe the  preparation of the pn outFOV data set. The method we adopt is designed to minimize the contamination from photons and SP shown in Section \ref{cont_pre}.\\
We retrieved Observation Data Files (ODF) from the {\it XMM-Newton} repository, then we performed a standard reduction using the 
{\it XMM-Newton} Science Analysis Software (SAS) v.14.0 and the calibration files available on 2016.
We barycentered all the data sets using the SAS tool {\tt barycen}.

\subsection{Data Set} \label{data_set}

The starting point of our project is the database of {\it XMM-Newton} observations produced within the EXTraS project \citep{del16}.
EXTraS is based on the 4$^{th}$ data release of the 3XMM source catalogue\footnote{http://xmmssc-www.star.le.ac.uk/Catalogue/3XMM-DR4/},
which considered all the {\it XMM-Newton} EPIC exposures made between 2000 February 3 and 2012 December 8.
For this analysis, we rely on the 5708 pn exposures (for a total time of $\sim$150 Ms) with an available  EXTraS background light curve.
We made the following sub-selections:\\
a. We used only observations performed in Full Frame or Extended Full Frame modes, because in the other sub-modes the outFOV area is not read;\\
b. We used only exposures with an attitude stability better than 5$''$, as reported in attitude files;\\
c. We excluded exposures taken, even partially, during solar energetic particle events (SEP), as defined by the ESA Solar Energetic Particle Environment Modelling application server\footnote{http://www.sepem.eu/} \citep[see Figure 10 of][]{gas17}, since SEPs with a particularly hard particle spectrum could in principle affect the outFOV rate on short time-scales;\\
d. We excluded highly SP-contaminated exposures and revolutions in order to minimize the contamination of the outFOV area (for details see Appendix \ref{off_script}).

\begin{figure*}
  \includegraphics[width=0.75\textwidth]{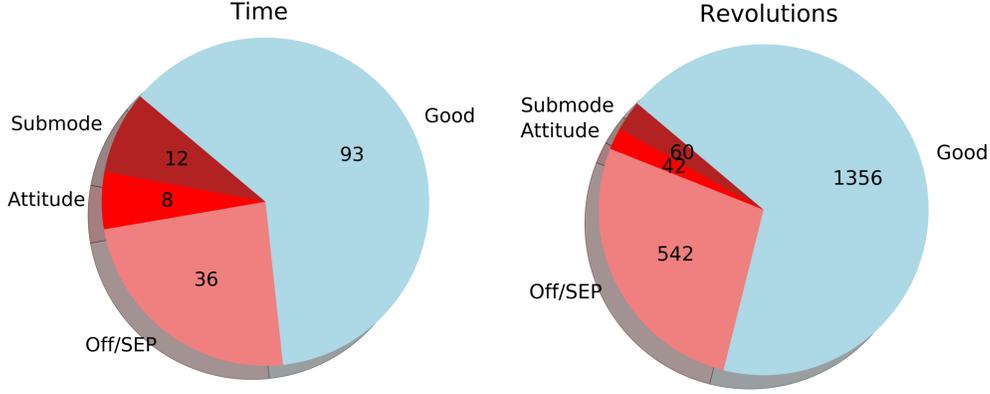}
  \caption{Exposure time excluded from the sample because of the different selections applied (as defined in Section \ref{data}) in Ms and in number of revolutions in the left and right panel, respectively. We show the exclusion due to submodes in which the outFOV area is not read (dark red), due to attitude instability (red) and due to the presence of SEP or proton flares (light red).}
\label{fig_data}
\end{figure*}

These selections lead to a final data set thet we dub "Main Data Set". It comprises $\sim$93 Ms, covering 1356 {\it XMM-Newton} revolutions.
In Figure \ref{fig_data}, we display the amounts of exposure time and revolutions excluded with respect to the initial EXTraS-based data set.

\subsection{Event Selection} \label{evt_sel}

We optimized our event selection to minimize the detector background and the contaminating SP and photons.
To exclude the contamination of the borders due to the pixellation of the detector, we define the outFOV region as the one
outside a 905"-radius circle centered in the geometrical center of the pn detector, while the inFOV region as the one within the co-centered 900"-radius circle.
For both regions, we excluded pixels close to CCD borders.

Following the standard analysis procedure, we applied a filter on event patterns, using only single and double events.
We used standard flags to exclude events with invalid patterns, associated to cosmic rays or MIPs, secondary and trailing events, events falling on and close to dead or bad pixels, bright pixels and bright columns.
Bright pixels and columns are not the same for all exposures, since the radiation damage increases with the instrument life time, and some of them are not excluded using standard flags. Instead of considering
separately each exposure, we chose to collect events from our entire data set and produce a total image, in detector coordinates. Then, we selected manually the remaining bright or dead regions and we excluded them in all exposures. 

To minimize the photon contamination, we use the energy band 10-14 keV, since the effective area for photons is negligible in this energy band.
To minimize the SP contamination in the outFOV we also perform a time selection to exclude SP flares. We perform this filtering applying for each exposure the algorithm developed for the EXTraS project on the inFOV data as described in Appendix \ref{off_script}.

All the expressions used to filter events through SAS are reported in Appendix \ref{app:SAS}.
Using these selections, we obtain an inFOV area of 28.355 cm$^2$, corresponding to 595.646 arcmin$^2$, and an outFOV area of 2.512 cm$^2$, corresponding to 52.780 arcmin$^2$. These define the inFOV and outFOV regions we shall use throughout this paper.

\section{Time Series: Extraction and Analysis} \label{data_analysis}

To obtain the surface brightness $sb$ (in units of counts
s$^{-1}$ cm$^{-2}$ keV$^{-1}$) of the outFOV area of pn, measured in the 10-14 keV energy band,
we started from our Main Data set as defined in Section \ref{data} and we extracted events in the outFOV region of each CCD using the SAS tool {\tt evselect}, with the selections reported in Section \ref{evt_sel}. The mean count rate in the outFOV region of pn is 0.053 counts s$^{-1}$, similar to the one obtained for the MOS2 outFOV: 0.071 counts s$^{-1}$ in the 7--9.4 \& 10--11 keV \citep{mar17}.
Then, we performed a few corrections and rescalings that we describe in the next paragraphs.  More specifically, we made a correction to exclude OoT events (Section \ref{oot_rescale}), we rescaled the raw number of events $c_j$ (where $j$ corresponds to the $j$-th CCD of the pn instrument) for the the exposed area (Section \ref{area_rescale}), for the active time (Section \ref{time_rescale}) and for the energy band (Section \ref{lc_creation}).

For the MOS2 camera, we studied the unconcentrated background on a ks time-scale \citep{gas17}. This is not possible for the pn camera since the exclusion of the time periods in which the inFOV background is high leads to a much lower statistics.
We shall therefore combine the obtained count rate of each CCD into a global pn count rate on timescales of exposures and orbit -- $\sim$ tens of ks (Section \ref{lc_creation}).
By investigating a sample of about 100 long MOS2 observations within the AREMBES/EXTraS data set our team was able to put a 3$\sigma$ upper limit of 5-8\%
on any residual fractional variability of the unfocused background on timescales shorter than 100 ks \citep[see][Gastaldello et al. in preparation]{vonkienlin18}.
We thus can assume that also the pn count-rate evaluated over an entire orbit is a good proxy for each moment within that orbit.

\subsection{Out Of Time Events Evaluation} \label{oot_rescale}

Events registered during readout (OoT events)
are assigned an incorrect positioning along the readout column. This results in a contamination in the outFOV area due to inFOV OoT events. In the following paragraphs we will describe the method we used to evaluate this component.

The task {\tt epproc} offers the possibility to simulate OoT events based on the original event list. By re-running the task, it is possible to create an event list
that treats all events as OoT events: the Y chip coordinate (RAWY) is simulated by randomly shifting the events along the RAWY axis. We extract events $c^{OoT}_{j,raw}$
from this list using the same prescriptions as in Section \ref{evt_sel}.
The number of resulting events must be multiplied for the ratio of the readout and integration time ($T_R/T_I$):
0.063 for Full Frame mode exposures and 0.023 for Extended Full Frame mode.
The current implementation of {\tt epevents} (called in the {\tt epproc} task)
does not automatically detect the instrument and bad pixel setting, thus a rescale of about 95\% must be applied manually\footnote{https://xmm-tools.cosmos.esa.int/external/sas/current/doc/epchain/index.html}.This is repeated for each $j^{th}$ CCD (where $j$ runs from 1 to 12).

Therefore, the number of OoT events expected in the outFOV area is:

\begin{equation}
c^{OoT}_j = c^{OoT}_{j,raw} \frac{T_R}{T_I} 0.95
\end{equation}

\subsection{Area Rescale} \label{area_rescale}

Standard SAS tools do not allow for exposure map creation outside the FoV. In order to rescale the counts detected from the outFOV area to the inFOV area, we relied on an outFOV cheesed mask $M_{out}$.
$M_{out}$ is produced following the outFOV definition reported in Section \ref{evt_sel}, covers 2.512 cm$^2$ (corresponding to 52.780 arcmin$^2$) and the size of its pixels is 1$"$.
For each $j^{th}$ CCD
we define $N_{j,out}$ as the number of pixels having a value equal to 1 in $M_{out}$ in the $j^{th}$ CCD.
For each CCD the area rescale factor $R_j$, measured in cm$^{-2}$, is defined as:

\begin{equation}
R_j = \frac{N_{j,out}}{2.512}
\end{equation}

\subsection{Good Time Computation} \label{time_rescale}

For each CCD we need to know the time in which the pn was taking events.
To compute the observing time for a single exposure and for the j$^{th}$ CCD, $T_j$, we must take into account different effects: the nominal Good Time Intervals (GTIs), the excluded on-flare time periods and the losses of exposure time.

The GTI extension of the event file provides the nominal lists $L_{j,GTI}$ of time periods in which each CCD is operating correctly.\\
These nominal lists of GTIs must be excised of on-flares periods.
Following the analysis reported in Appendix \ref{off_script}, we derived the list $L_{on}$ of on-flares times of each observation.
By removing these intervals from each $L_{j,GTI}$ we got for each CCD the lists of GTIs of off-flares periods $L_{j,off}$.\\
Discarded CCD columns and rejection of MIPs are registered as losses of exposure times.
Such losses of exposure times (together with the dead time computation) are accounted for by renormalizing the nominal frame time $T_f$ of each time frame
through the FRACEXP column in EXPOSU extension of the event file, which provides the fractional exposure, $f_i$, of each time frame $i$.
The effective observing exposure of each time frame $i$ is $T_i^{eff}= T_f *f_i$.
The total good time $T_j$ for each CCD is the sum of the effective exposures on all the $n$ $i^{th}$ time frames in the CCD $L_{j,off}$ list:

\begin{equation}
T_j = \sum_{i=1}^{n}{T_i^{eff}}
\end{equation} \label{eq_time}

\subsection{Time Series Computation} \label{lc_creation}

\begin{figure*}
  \includegraphics[width=0.8\textwidth]{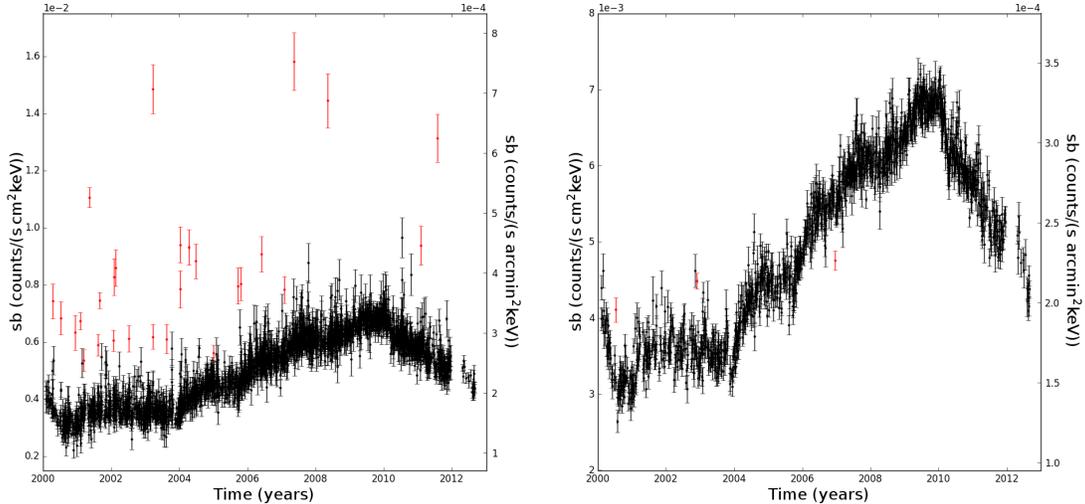}
  \caption{Surface brightness extracted from outFOV area of pn ($sb$, Section \ref{lc_creation}) as a function of the observing date. Red points are the ones excluded from the selection defined in Section \ref{out}. {\it Left panel:} We report the values of $sb$ evaluated during the time of single exposures, computed using Equation 4. {\it Right panel:} We report the values of $sb$ evaluated during the time of single revolutions.}
\label{fig_results}
\end{figure*}

For each exposure and for each revolution, we computed the surface brightess $sb$ and its error $\sigma_{sb}$ in the outFOV region, measured in the in the 10-14 keV energy band, as:\\
\begin{equation}
sb = \frac{1}{\Delta E}\sum_{j=1}^{12}{\frac{c_{j}-c^{OoT}_j}{T_{j}R_{j}}} \quad ,\quad \sigma_{sb} = \frac{1}{\Delta E}\sqrt{\sum_{j=1}^{12}{\frac{c_{j}+c^{OoT}_{j}}{T_{j}^2R_{j}^2}}}
\end{equation} \label{eq_lc}

\noindent where $c_{j}$ (Section \ref{data_analysis}) is the number of observed counts, $c^{OoT}_j$ (Section \ref{oot_rescale}) is the predicted number of OoT counts,
$R_j$ (Section \ref{area_rescale}) is the area rescale factor, $T_j$ (Section \ref{time_rescale}) is the total good time
and $\Delta E=4$ keV is the energy band width.\\
The surface brightness time series by exposure and revolution are reported in Figure \ref{fig_results}. Using Equation 4 to evaluate the surface brightness of our entire sample we obtain
$sb_{sample}=(5.118\pm0.003)\times10^{-3}$ counts s$^{-1}$ cm$^{-2}$ keV$^{-1}$.

\subsection{Outliers} \label{out}

\begin{figure*}
  \includegraphics[width=0.75\textwidth]{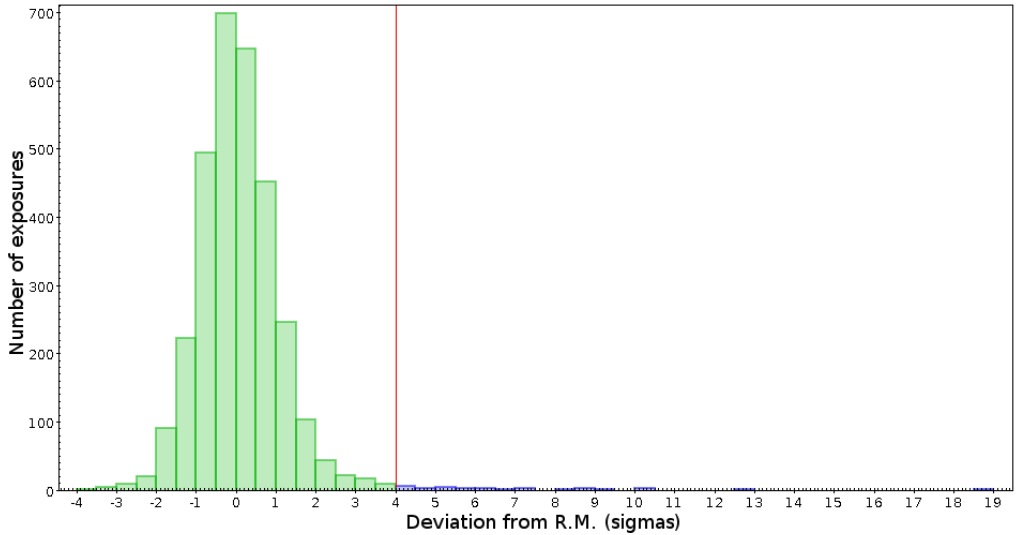}
  \caption{Histogram of deviations, in $\sigma$s, of the $sb$ value computed for each exposure from the result of the running mean analysis (Section \ref{out}).
  Exposures with deviations larger than $\pm4\sigma$ (red line) are identified as outliers and shown in blue.}
\label{fig_RM}
\end{figure*}

From the analysis of the time series reported in Figure \ref{fig_results}, it is apparent that there are some exposures which are clear outliers to the general behaviour of our data set. In order to identify them in an objective way, we calculated the running mean on 4 nearby data points and study the distributions of deviations from the mean (Figure \ref{fig_RM}).
From Figure \ref{fig_RM}, it is apparent that the tails of the distribution of deviations are highly asymmetric:
there are 28 outlier exposures (over 3104, about 0.9\%) that show an excess $>4\sigma$ (marked in red in the left panel of Figure \ref{fig_results}), while no exposure deviates at more than $-4\sigma$.
This cannot be due to a statistical effect, and thus we expect that these ``excesses'' contain some form of contamination.

To investigate this issue, we inspected the EXTraS light curves, the Radiation Monitor\footnote{https://www.cosmos.esa.int/web/xmm-newton/radmon-details} light curves and pn images of these exposures.\\
We noticed that in 18 of the 28 observations the off-flare exposure time was lower than 3 ks. We suspect that these exposures are entirely contaminated by SP flares. Under such circumstances, the cleaning procedure described in Appendix \ref{off_script} can incorrectly assume the four bins with lowest count-rate to be flare free.\\
Furthermore In 12 of the 28 exposures the Radiation monitor light  curve suggests the presence of low residual levels of SEPs
not recognized by the ESA Solar Energetic Particle Environment Modelling application server \citep[see e.g. Figure 5 from ][]{gas17}. Since we excluded SEP periods, these exposures should be excluded from our sample.\\
Finally, in 7 of the 28 exposures we found hard, bright sources falling on the edge of FoV. Thus, we cannot exclude a photon contamination of the outFOV coming from such sources.\\
Since it is apparent that the selected exposures have an outFOV highly contaminated by protons, contaminated by photons and/or fall during SEP periods, we will conservatively mark them as outliers. They will be treated separately (where possible, otherwise excluded) in the following analysis. We shall make a posteriori considerations on these outliers
in Section \ref{pnmos2}.

After the removal of the 28 excesses, we repeated the computation of $sb$ as in Equation 4 for each revolution, instead than for each exposure.
Again, a running mean analysis revealed three revolutions (over 1356) deviating for more than $\pm4\sigma$ (marked in red in the right panel of Figure \ref{fig_results}); these will be treated separately (where possible, otherwise excluded) in the following analysis. 

\section{Time Series: Results} \label{filtermode}
\subsection{Analysis of different optical filters and different sub-modes} \label{filmode}

\begin{figure*}[hbt!]
  \includegraphics[width=0.75\textwidth]{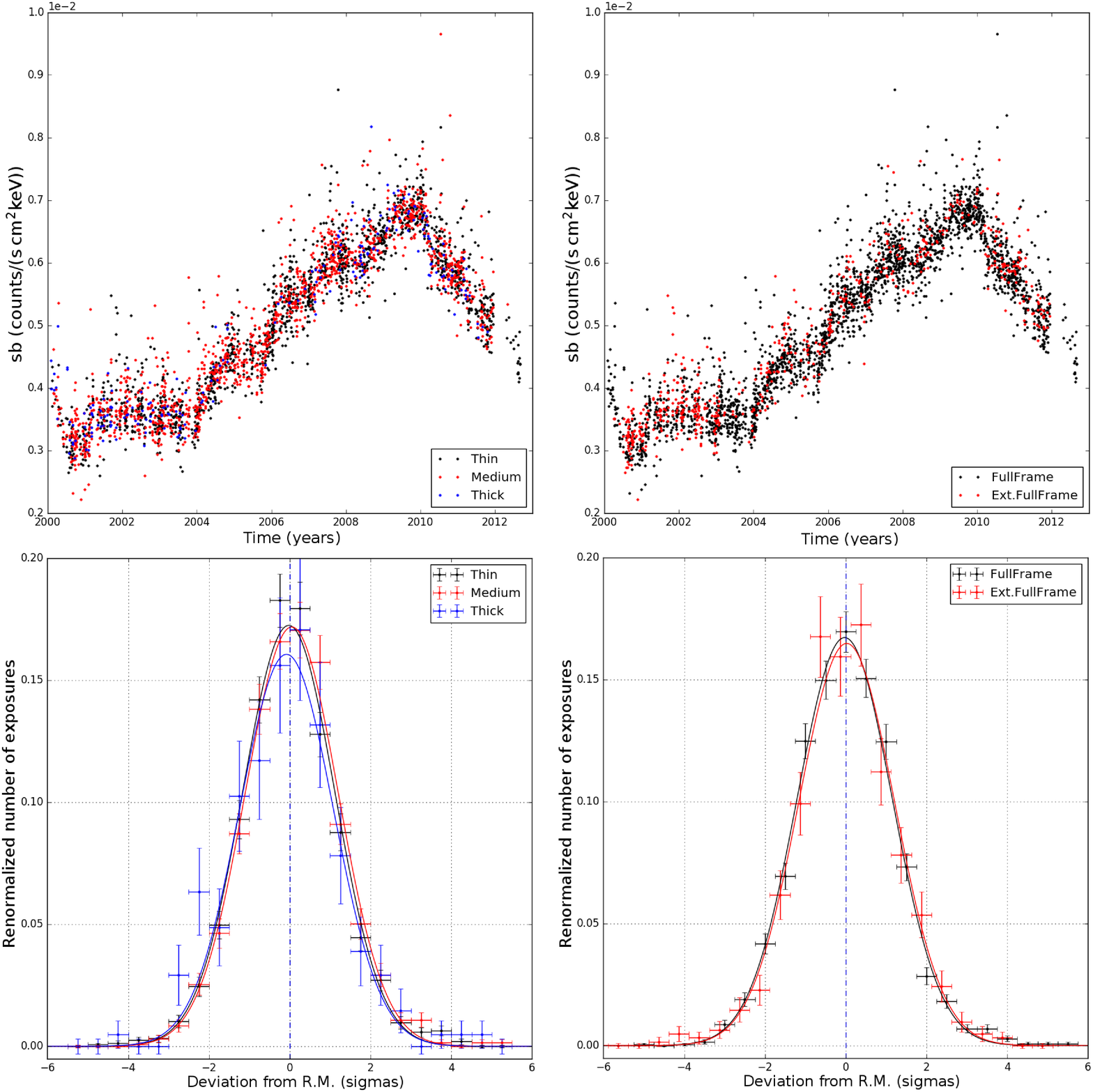}
  \caption{{\it Upper Panel:}Surface brightness extracted from outFOV area of pn (Section \ref{lc_creation}) against time for individual exposures and computed from Equation 4. {\it Left panel:} Exposures taken with Thin optical filter are marked in black, with Medium optical filter in red and with Thick optical filter in blue. {\it Right panel:} Exposures taken with the FullFrame submode are marked in black and with the ExtendedFullFrame submode in red.\\
  {\it Lower Panel}: Histograms of deviations, in $\sigma$s, of the $sb$ value computed for each exposure of a given filter/mode from the result of the running mean analysis performed on the other filters/mode (Section \ref{filmode}), and the best-fitting Gaussian. Colours as in the upper panel.}
\label{fig_filtermode}
\end{figure*}

In order to test if different observation settings -- such as the choice of the optical filter or the observing sub-mode -- can change the outFOV surface brightness, we will analyze the time series of outFOV $sb$ separating exposures performed with different optical filters and different submodes.

Each EPIC camera is equipped with a set of three separate optical filters, named thick, medium, and thin. Although these filters absorb part of the incoming focused events (both from photons and soft protons), they are almost transparent to high-energy particles and therefore $sb$ should not depend on the filter used. Figure \ref{fig_filtermode} (left panel) reports the outFOV surface brightness as a function of time for different optical filters.The three samples of surface brightness appear to be consistent.\\
In order to  test statistically the differences among these distributions, we must consider that there they have a different occurrence, e.g. the thick filter was used more often during the first years of operations. This, coupled with the variation of the surface brightness with time, requires a binning (on the exposure-based data set) to perform most of the statistical tests. In analogy with the algorithm used to estimate the outlier exposures and revolutions (Section \ref{out}), we compared the deviations between the surface brightness for a given filter against the Running Mean calculated on the four nearby data points with a different filter. As we did in Figure \ref{fig_RM}, we built histograms (one for each filter) of such deviations; then we normalized for the total number of exposures and we plotted a Gaussian model (see Figure \ref{fig_filtermode}, lower panel). A difference in the overall population would result in a difference in the distributions. We found that the parameters are consistent within 1$\sigma$ (see Table 1).\\
We tried a different approach, binning in time the exposures with a step of one month, and we tested the three binned distributions.
We performed both a Kolmogorov-Smirnov test -- using the binned surface brightness -- and a multidimensional Kolmogorov-Smirnov test -- using both time and surface brightness -- on each pair of the three
distributions \citep{pea83,fas87,pre92}. The three distributions are consistent within 5$\sigma$.\\
As a further test we plotted the binned $sb_{thin}$, $sb_{medium}$ and $sb_{thick}$ obtained with the one-month step in pairs; for each pair, the points are best fitted by a linear distribution compatible, within 3$\sigma$, with a 1:1 relation.

We also tested for  differences associated to the 2 different sub-modes, Full Frame and Extended Full Frame, which differ for time resolution
(73.4 ms and 199.1 ms, respectively) and the ratio of integration and readout time (6.3\% and 2.3\%, respectively), leading to a different percentage of OoT events.
If OoT events are treated correctly, we do not expect any difference
between the two samples of exposures. The two samples are showed in Figure \ref{fig_filtermode} (right panel).
We followed the same procedure as before to group and test the two distributions.
Again, the two distributions are consistent using the Running Mean method (within 1$\sigma$) and using the Kolmogorov-Smirnov tests (within 5$\sigma$), and the linear fit of $sb_{FF}$ versus $sb_{EFF}$ is compatible (within 3$\sigma$) with a 1:1 relation.

We thus found no variation of the outFOV surface brightness that we extracted either with the sub-mode or with the optical filter used.
While the first result is useful only as a test of the method we used, the latter result provides an important clue about the particles responsible for the outFOV count rate. This will be discussed in Section \ref{disc}. 

\begin{table}
\begin{tabular}{lccc} \hline
Filter/Mode & Maximum & Maximum position & $\sigma$ \\
 & & & \\ \hline
F.F. & 0.167$\pm$0.006 & -0.03$\pm$0.03 & 1.18$\pm$0.02\\
E.F.F. & 0.165$\pm$0.007 & 0.02$\pm$0.04 & 1.18$\pm$0.03\\ \hline
Thin & 0.172$\pm$0.006 & -0.02$\pm$0.03 & 1.14$\pm$0.03\\
Medium & 0.172$\pm$0.005 & 0.05$\pm$0.03 & 1.15$\pm$0.02\\
Thick & 0.161$\pm$0.012 & -0.08$\pm$0.08 & 1.16$\pm$0.06\\ \hline
\end{tabular}
\caption{Parameters of the best-fitting Gaussians of the histograms of deviations, in $\sigma$s, of the $sb$ value computed for each exposure of a given filter/mode from the result of the running mean analysis performed on the other filters/mode (Section \ref{filmode} and Figure \ref{fig_filtermode}. We report the value of the maximum, the maximum position and the $\sigma$ of each Gaussian.)}
\end{table}

\subsection{pn-MOS2 comparison} \label{pnmos2}

The pn camera is composed by twelve back-illuminated pn-CCDs, while the MOS cameras are composed by seven EEV type 22 front-illuminated CCDs. They differ in sensitivity thickness, quantum efficiency and energy response \citep[see e.g.][]{str01,tur01}. A comparison among pn and MOS cameras thus offers a
unique opportunity to analyze the sensitivity to unconcentrated background
of different X-ray detectors that share the same particle environment.\\
To perform this we rely on the data set of MOS2 outFOV count rates reported in \citet{mar17}. Since we do not have an analogous data set for the MOS1 camera \citep[see also the discussion in][]{mar17}, we restrict our comparison to MOS2.
From the starting MOS2 data set we excluded time periods at the beginning and end of each orbit (orbital phase $<$0.15 or $>$0.8), since the outFOV is possibly contaminated by the Earth radiation belts \citep{kun08}.\footnote{We checked the importance of such selection for pn: since most of these periods are already excluded by the high-background periods cut, all the results reported in this paper do not change.} 
Then, we performed the same analysis of outlier exposures as for pn (Section \ref{lc_creation}),
flagging 34 over 7493 MOS2 exposures ($\sim$0.5\%). Similarly, we flagged 32 over 2395, $\sim$1\%, revolutions as outliers.\\

In order to compare directly the response of the two cameras, we normalized MOS2 rates taken from \citet{mar17} for the area. We also took into account the different energy bands that we used:
for MOS2 we fitted the spectrum of outFOV events \citep[taken from][]{mar17} using a diagonal response matrix -- appropriate for particle-induced events -- we extrapolated the rate in the
10-14 keV energy band width yielding a corrective factor of 2.682 keV$^{-1}$.

\begin{figure*}
  \includegraphics[width=0.75\textwidth]{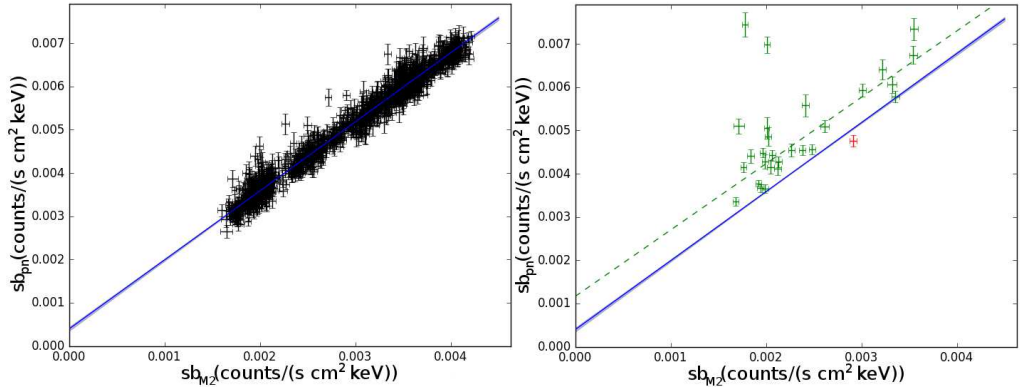}
  \caption{Revolution-based surface brightness for pn ($sb$, Section \ref{pnmos2}) versus MOS2. The best fitting linear relation obtained in Section \ref{pnmos2} is shown in blue, with 1$\sigma$ uncertainties in light blue. {\it Left panel:} We report the revolution-based $sb$ obtained by exceeding outlier exposures and revolution (as defined in Section \ref{out}). {\it Right panel:} In green, we show the revolutions that are outliers, or that contain outlier exposures, in the pn camera distribution (in green). In red we mark the only revolution that is an outlier for both pn and MOS2. The dashed-green line reports the best fit of these outlier points.}
\label{fig_pnm2}
\end{figure*}

We considered only revolutions that are covered both by pn and MOS2 data sets. We note that events extracted from MOS2 and pn usually do not cover the
same time period due to the different GTI selection and pn off-flare selection (which was not necessary in the MOS2 corner). Tipically, MOS2 covers
a longer time period than pn, or the considered time periods can be few ks distant. As discussed in \citet{vonkienlin18}, MOS2 outFOV count rate usually does not show evidence of variability on orbit time scales: we shall therefore proceed with our pn vs MOS comparison without making any correction for these small differences (which will be seen as a spread in the distribution).
Figure \ref{fig_pnm2} reports the relation among the surface brightness per energy of the pn and MOS2 cameras ($sb_{\mathrm{pn}}$ and $sb_{\mathrm{M2}}$ respectively);
pn outlier exposures or revolutions are considered separately (right panel).
In order to perform the fit, we relied on the {\tt ltsfit} python module that performs a robust linear fit to data including errors on both variables, while allowing for possible intrinsic scatter \citep[][Section 3.2]{cap13}. We obtain a clear correlation that follows:\\
\begin{equation}
sb_{\mathrm{pn}} = \alpha (sb_{\mathrm{M2}}-sb_{\mathrm{mid}})+ \beta
\end{equation} \label{eq_pnmos2}
where the inclination is $\alpha$=1.596$\pm$0.006 (1$\sigma$ error), the constant is $\beta$=(5.240$\pm$0.005)$\times10^{-3}$ counts s$^{-1}$ cm$^{-2}$ keV$^{-1}$ and
$sb_{\mathrm{mid}}$=3.035$\times10^{-3}$ counts s$^{-1}$ cm$^{-2}$ keV$^{-1}$ is the pivot point. The relation intersects the y axis (MOS2 brightness null) at
$sb_{pn,0}=(3.9\pm0.4$)$\times10^{-4}$ counts s$^{-1}$ cm$^{-2}$ keV$^{-1}$.

The intrinsic scatter on the y axis is $sb_{sc}$=1.17$\times10^{-4}$ counts s$^{-1}$ cm$^{-2}$ keV$^{-1}$ (2.3\% of the surface brightness evaluated over our entire sample, $sb_{sample}$, see Section \ref{lc_creation}).

Although the 28 outlier pn exposures (Section \ref{out}) have not been fitted, they do not change significantly the results of the fit. They are mostly above the fitted relation: the presence of an excess of counts in the pn outFOV undetected by MOS2 confirms that these exposures are exceptionally contaminated in the pn camera, probably due to SP and photons.

\section{Imaging} \label{imaging}

\begin{figure*}
  \includegraphics[width=0.75\textwidth]{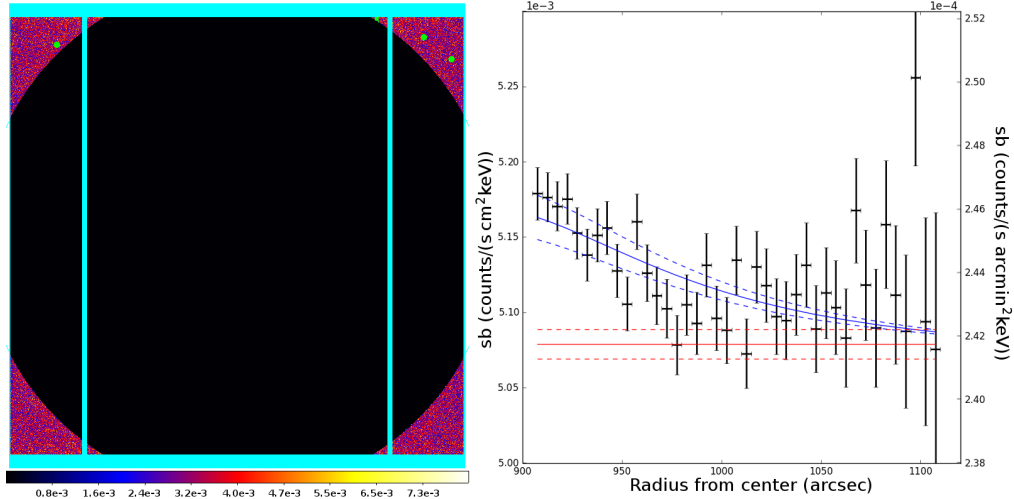}
  \caption{{\it Left Panel:} Surface brightness map defined as in Section \ref{imaging}, for our entire sample, in detector coordinates and units of counts s$^{-1}$ cm$^{-2}$ keV$^{-1}$.
  We show in cyan the CCD borders and nearby excluded pixels.
  In black we mark the area excluded from the outFOV definition. In green we mark the area excluded for bad pixels and columns.
  {\it Right Panel:} Surface brightness profile from the geometrical center of pn detector with a $5"$ step, computed following Equation 6. We report the best fit with a constant model plus a Lorentzian, shown in red and blue, respectively, with 1$\sigma$ errors, using the model in Equation B.1 (Appendix \ref{app:cont}) and prescriptions described in Section \ref{imaging}. The difference between the integral of the blue component and the integral of the red component, renormalized for the area, gives the contamination due to concentrated photons and soft protons.
}
\label{fig_bp}
\end{figure*}

To study the spatial distribution of the unconcentrated background in the outFOV area, we produced the total count map $C_{out}$, by summing the images of all exposures in our data set and with our event selection. 
All images are in detector coordinates and have a pixel size of 1$"$.
Similarly to $C_{out}$, we created the total OoT count map $C^{OoT}_{out}$, corrected as in Section \ref{oot_rescale}.
We produced the total exposure map $T$ by evaluating, for each pixel in the outFOV area, the exposure time following equation 3.
Figure \ref{fig_bp} (left panel) reports the surface brightness map, defined as $(C_{out}-C^{OoT}_{out})/T$.
A first inspection of the surface brightness map does not reveal any spatial dependence of the unconcentrated background. As a further test, we analised the radial surface brightness profile.

The radial profile could be affected by systematic effects induced by pixelisation.
Indeed, the event selection described in Section \ref{evt_sel} was performed with SAS expressions (in principle without any pixelisation),
while we applied an area rescale based on the 1 arcsec-pixel mask $M_{out}$ (Section \ref{area_rescale}). 
This effect causes an artificial lower count rate at the borders of the mask and/or some counts falling just outside the mask borders.
We estimate this effect to cause a count rate loss that artificially lowers the $sb$ from equation 4 of $\sim$0.2\% (evaluated over the entire data set). This is about an order of magnitude smaller than the intrinsic scatter on the relations reported in Section \ref{pnmos2} and therefore negligible.
Surface brightness profiles would instead be distorted since this effect is located on the border pixels.
For the imaging analysis, we shall therefore exclude the border pixels.

In order to produce a radial brightness profile, we used annuli centered on the geometrical center of the pn camera and with width $v=5"$.
For each annulus centered at radius $r$
we extracted the area-corrected observed counts as $c(r)=\sum_{i=1}^{n}{C_{out,i}\bar{M}_{out,i}}$, where $i$ is an image pixel
over the $n$ pixels within the annulus and $\bar{M}_{out,i}$ is a mask modified to exclude border pixels. Similarly, we extracted the area-corrected OoT counts $c^{OoT}(r)$ (as in Section \ref{oot_rescale}).
For each annulus, we extracted the exposure time as $t(r)=\sum_{i=1}^{n}{T_i \bar{M}_{out,i}}$.
We introduced a conversion factor on the area to convert from pixel$^2$ (where 1 pixel = 1 arcsec) to cm$^2$: $f_{A}=1.322\times10^{5}$ cm$^2$ pixel$^{-2}$.\\
We built the surface brightness radial profile using, for each annulus:

\begin{equation}
sb(r) = \frac{c(r)-c^{OoT}(r)}{t(r)f_{A}}
\end{equation} \label{eq_brig}

\noindent The resulting surface brightness profile is reported in the right panel of Figure \ref{fig_bp}.

The radial surface brightness profile in the outFOV region features a very modest but statistically significant decline: a fit with a constant model results in a null hypothesis probability of $3.1\times10^{-4}$.
A small excess is present in the inner region of outFOV, likely associated to residual contaminants (concentrated SP and photons).
To provide a realistic model of the radial profile of the contamination in the outFOV region, we decided to use the data set of contaminated periods, obtained during SP flares, since it provides a much larger statistics. While the details of the analysis are described in Appendix \ref{app:cont}, we briefly report here that we modeled the radial profile with an empirical model composed by a Lorentzian function plus a constant (Equation B.1), whose parameters (except for the normalization) do not depend on the intensity of the flares. We thus apply this model to our profile in Fig. 7, fixing the width and the maximum of the Lorentzian to the best fit values obtained in Appendix \ref{app:cont} ($w=113.4''$ and $r_0=875.6''$, respectively).
Then we fit the profile of our data set with this model, leaving as free parameters the normalization and the constant.
In this way we obtain an uncontaminated brightness of $sb_{real} = (5.08\pm0.01)\times10^{-3}$ counts s$^{-1}$ cm$^{-2}$ keV$^{-1}$. Thus, the contamination accounts for less than 1\% of the unconcentrated background surface brightness evaluated over our sample ($sb_{sample}$, Section \ref{lc_creation}).

To evaluate the spatial distribution of counts within the outFOV, we extracted the surface brightness of each quadrant. We found that quadrants 2 and 3 (the rightmost quadrants in detector coordinates) are in agreement within 1 $\sigma$, as well as quadrants 1 and 4. Quadrants 2 and 3 have a surface brightness significantly ($>5\sigma$) lower by $\sim$2\% than the one of quadrants 1 and 4. We conclude that the unconcentrated background extracted from the outFOV region in the 10--14 keV energy band does not show spatial variations higher than 2\%.

\section{Discussion} \label{disc}

We have presented clear and incontrovertible evidence that corners of the pn detector are exposed to photons and particles concentrated by the X-ray telescope (see Section \ref{cont_pre}).
This contamination features a radial profile, well modeled by a quasi-Lorentzian function, which varies in intensity but not in shape (see Appendix \ref{app:cont}), it covers the entire outFOV, and thus  prevents us from directly using the corner data to measure the unconcentrated particle background, as done for the MOS. We therefore devised a modified approach to assess this component consisting in: first cleaning corner data from photon and proton contamination as best as possible and then measuring the background rate from the outFOV area.\\
The validity of this method is confirmed by the analysis of the radial brightness profile of our entire sample (Section \ref{imaging}) which revealed a total residual contamination of less than 1\% of the total surface brightness. Moreover, the low scatter in the relation between outFOV surface brightness of pn and MOS2, with the latter unaffected by this contamination, has shown that the difference in the level of contamination between different revolutions is low. Indeed this must be included in the intrinsic scatter in the pn vs MOS2 correlation -- 2.3\% of the surface brigthness evaluated over our sample, $sb_{sample}$. This can be adopted as an upper limit to any difference in contamination between different revolutions.

Having devised for the first time a technique to use pn corner data to monitor the unconcentrated particle background, we now go on to discuss: 1) effective proxies of the pn background and 2) the implications on the nature of the pn particle background.

\subsection{Proxies of the pn Background} \label{disc_proxies}

To evaluate the unconcentrated background in the inFOV, a first possibility is to use the corner data of the pn. As shown in Section \ref{cont_pre} and Appendix \ref{app:cont}, the presence of a substantial contamination from concentrated SP and photons is a clear limitation to this approach. The data needs to be cleaned carefully for soft proton flares before a measure of the pn background from the corner can be made. Note that this method works less than perfectly on observations where soft proton cleaned data is insufficient to allow a measurement. A large fraction of the outliers in Figures \ref{fig_results} and \ref{fig_RM} are probably due to this effect.
Finally, since concentrated photons contribute to the contamination of the outFOV region, such method should be avoided in case of bright, extended celestial sources detected in the inFOV. 

A second possibility to evaluate the unconcentrated background in the inFOV is to measure the unconcentrated background from MOS2 corner data, following the method described in \citet{mar17}, and estimate the pn unconcentrated background in the outFOV from the relationship between MOS2 and pn presented in Section \ref{pnmos2}. The resulting pn surface brightness should take into account a systematic error of 1.17$\times10^{-4}$ counts s$^{-1}$ cm$^{-2}$ keV$^{-1}$, which reflects the intrinsic scatter in the MOS2/pn relation. This surface brightness can be extended to the entire pn energy band using the unconcentrated background spectral modelling derived from closed data or blank sky fields \citep[see e.g.][]{kat04,fre04,fra14}.\\
This option has the advantage that no soft proton flare cleaning needs to be performed since MOS2 corners are properly shielded from the sky. It also affords better statistics and temporal coverage than what is available through the pn corner data. As discussed in Section \ref{pnmos2}, the non-simultaneity between MOS2 and pn measurements, as well as any other effect, induce systematic errors that are contained within the 2.3\% intrinsic scatter measured in the MOS2 vs pn correlation.

A third alternative which has not been explored in this paper consists in the use of pn "discarded lines".
A column is discarded when an event above the upper limit threshold is detected: this procedure is called "minimum ionizing particle" (MIP) rejection. Since  the vast majority of the MIP events are particle events, the number of discarded columns is tightly related to the unfocussed background count rate.
This option has been investigated in the past within the EPIC pn team \citep[see e.g.][]{kat04,fre04} and more recently by the SOC using NDSLIN in EPIC-pn as a proxy for the QPB (I. de la Calle, XMM-Newton SOC, 2019\footnote{available at https://www.cosmos.esa.int/web/xmm-newton/background}).

\subsection{Origin of the pn Unconcentrated Background} \label{disc_origin}

\begin{figure*}[hbt!]
  \includegraphics[width=0.75\textwidth]{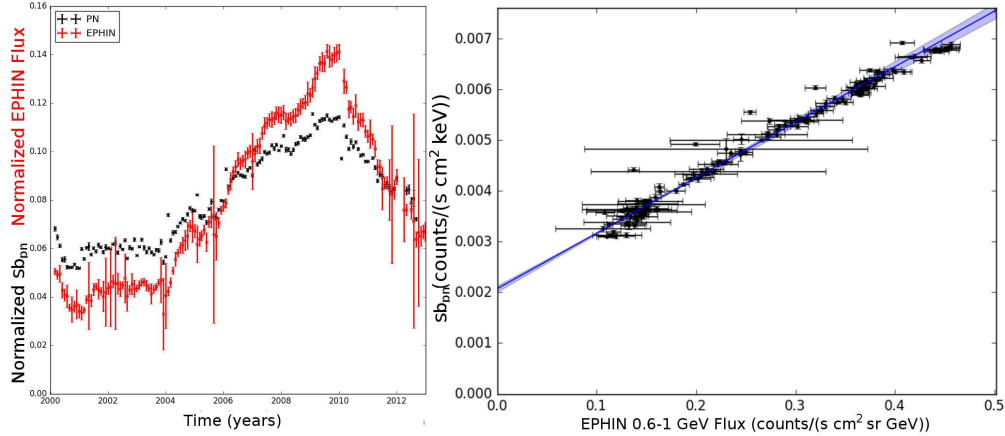}
  \caption{{\it Left Panel:} $sb_{\mathrm{pn}}$ (counts s$^{-1}$ cm$^{-2}$ keV$^{-1}$, black) and EPHIN flux (0.6-1 GeV band, counts s$^{-1}$ cm$^{-2}$ sr$^{-1}$ GeV$^{-1}$, red) versus time. We normalized $sb_{\mathrm{pn}}$ and EPHIN flux to ease comparison between the two time series. EPHIN error bars take into account both statistical and systematic error \citep[see Section 3 of][]{kul16}
    {\it Right Panel:} $sb_{\mathrm{pn}}$ versus EPHIN flux (0.6-1 GeV band). The best fit relation, with 1$\sigma$ errors and fitted using the same procedure as in Section \ref{pnmos2}, is shown in blue. It has the form $sb_{pn} = m* (F_{EPHIN}-F_{mid})  + q$, where $sb_{pn}$ is the pn surface brightness, $F_{EPHIN}$ the EPHIN flux, 
$q$ = (4.47$\pm$0.01) $\times 10^{-3}$ counts s$^{-1}$ cm$^{-2}$ keV$^{-1}$, is the constant term evaluated at the pivot point $F_{mid}= 0.245$ counts s$^{-1}$ cm$^{-2}$ sr$^{-1}$ GeV$^{-1}$ and $m$ = (1.09$\pm$0.01) $\times 10^{-2}$  (counts s$^{-1}$ cm$^{-2}$ keV$^{-1}$)/(counts s$^{-1}$ cm$^{-2}$ sr$^{-1}$ GeV$^{-1}$), the slope; the relation intersect the y axis ($F_{EPHIN}$ null) at $sb_o$ = 2.07$\pm$0.06 $\times 10^{-3}$ counts s$^{-1}$ cm$^{-2}$ keV$^{-1}$.}
\label{fig_ephin}
\end{figure*}

In this section we  investigate the origin of the pn background. To this aim, we  introduce a set of measurements of high energy protons collected with the EPHIN detector onboard SOHO, see \citet{kul16} for details. 

In Figure \ref{fig_ephin} (left panel) we show a time series (on a month time scale) for high energy protons in the range 630--970 MeV accumulated with EPHIN. To ease comparison between the two time series we normalized them setting their integrals equal to one.
Similarly we produced the time series of pn corners on a month time scale using Equation 4.
Comparison of EPHIN and pn time series reveals a striking similarity between the two. 
To investigate further, we correlated the EPHIN and pn time-series, finding a remarkably tight relationship between the two, see Figure \ref{fig_ephin} (right panel). A linear fit shows that, baring a few outliers, quite likely originating from the pn (see e.g. Section \ref{out}), the vast majority of data-points are clustered very tightly around the best fitting line. To evaluate this, we do not rely on the intrinsic scatter, as in Section \ref{pnmos2}, since it assumes that errors are statistical in nature and, in the case of EPHIN data, the systematic errors are important \citep[see Section 3 of][]{kul16}. We shall instead use the total scatter on the y axis, defined as:

\begin{equation}
s_t = \sqrt{\frac{\sum_{i=1}^{N}{(y_i-y(x_i))^2}}{N}} ,
\end{equation}

\noindent where $(x_i,y_i)$ are the $N$ fitted points (including the aforementioned outliers). $s_t$ is 1.66$\times10^{-4}$ counts s$^{-1}$ cm$^{-2}$ keV$^{-1}$, namely 3.5\% of the median of the pn monthly renormalized rate. Note that the total scatter includes the intrinsic scatter and can be regarded as an upper limit for the latter.
This comparison strongly suggests that cosmic ray protons are one of the main contributors to the unconcentrated pn background.
The lack of any detectable differences between unconcentrated background rates measures with different pn filters and sub-modes (see Section \ref{filtermode}) is consistent with this interpretation. Indeed, as shown by simulations carried out on instruments similar to the pn \citep{vonkienlin18}, the unconcentrated background associated to high energy protons is traced back to secondaries, mostly electrons and photons, produced  in the immediate vicinity of the detector and there is no reason to assume that either filter or sub-mode affect these processes.

Although the EPHIN and pn time-series bare a striking resemblance, they also differ in one important respect: the maximum to minimum ratio for the EPHIN curve is significantly larger than for the pn, (see Figure \ref{fig_ephin}). When performing a linear fit to the correlation between the two quantities, such a difference manifests itself as a non zero constant term. This is indeed what is found, the pn surface brightness extrapolated to 0 EPHIN rate is $sb_{\mathrm{o}}=(2.07\pm0.06)\times10^{-3}$ counts s$^{-1}$ cm$^{-2}$ keV$^{-1}$, $\sim$30\% to $\sim$70\% of the revolution-based surface brightnesses. 
The most obvious interpretation of this result is that there is some other agent contributing to the pn background beyond cosmic ray protons. Whatever the nature of this component, it must vary little or not at all over the solar cycle. Indeed, any variation uncorrelated with the solar cycle would lead to a scatter in the EPHIN vs pn relation. Thus the total scatter $s_t$ can be used to set an 8\% upper limit on any variation of $sb_{\mathrm{o}}$.
The question we ask ourselves is: what could be responsible for this component?  Following standard scientific practice, we shall start by investing the eventuality that it is some artifact in the data.
A possibility that we can quickly rule out is the existence of some spurious correlation between the EPHIN and pn data sets: the measures were extracted from detectors
located $\sim 1.5\times 10^6$ km apart and were reduced and analyzed independently of each other. Another possibility is calibration. \citet{kul16} have estimated a calibration uncertainty of about 20\% on EPHIN data, this uncertainty could be related to the absolute or relative calibration.
If it is related to the absolute calibration, than it has no impact on the constant term in the linear relation. Indeed, if the EPHIN fluxes were
all systematically off by a certain percentage (of the fluxes themselves) it would lead to a change in the slope of the correlation, not on the constant term (since the zero point of the EPHIN flux does not change). Conversely, if the issue were on the relative calibration, i.e. if values of single points were offset with respect to other data points in the EPHIN time series, we would observe a scatter in the relation with the pn detector. Since the scatter in the relation is small, this implies that any relative calibration issue must also be small, less than 1.52$\times10^{-2}$ counts s$^{-1}$ sr$^{-1}$ GeV$^{-1}$,
that is 6\% of the median on the EPHIN monthly fluxes. This is estimated from the best fit parameters and associated uncertainties.
Note that this argument can be turned around and applied to the pn data with similar results, any relative calibration uncertainty on pn fluxes must be contained within 3.5\% of the pn surface brightness evaluated over our entire sample $sb_{sample}$.

\subsubsection{The constant component} \label{const_comp}

Having found no evidence that the constant term is an artifact, we proceed with an evaluation of any systematic error that might be affecting it. Before reaching the pn detector Cosmic Ray (hereafter CR) protons must travel through a significant amount of matter, roughly 3 cm of equivalent Al \citep{hal07}. Under such conditions, lower energy protons,
which are of course more copious, stand a lower chance of traversing the shielding without being absorbed than higher energy protons. The net effect is that the protons responsible for the pn background have a broad distribution in initial energy, ranging from a few hundreds of MeV to several GeV. Since the solar modulation affecting CR protons depends upon their energy, with the more energetic protons less modulated \citep[see e.g.][]{pot17}, the modulation of the proton induced pn background will result from a weighted mean of the modulation of CR protons. If the distribution of matter around the pn detector were known rather precisely, we could compute an appropriately weighted EPHIN proton time series, correlate it with the pn time series and derive a robust estimate of the constant term. Given our lack of detailed knowledge of the pn camera, we are forced to take a more conservative approach: we repeat our estimate of the constant term by correlating the pn data with EPHIN proton curves of sufficiently low and high energies to derive an upper and lower bound to the constant term. To this aim, we used the curves of 500 MeV and 1.2 GeV . We find that in the first case the constant term is $\sim2.5\times10^{-3}$ counts s$^{-1}$ cm$^{-2}$ keV$^{-1}$ and in the latter case it is $\sim1.3\times10^{-3}$ counts s$^{-1}$ cm$^{-2}$ keV$^{-1}$.

Having derived a robust estimate of the constant component, as we shall henceforth call it, we come to the question of its origin.
A possibility we can discard rather quickly is that it comes from cosmic ray electrons or alpha particles. Since both these species follow a solar cycle modulation very similar to the one for protons, they cannot be responsible for a component that is consistent with being constant over the solar cycle.\footnote{The modulation could be slightly different for different species, so a very small part of the constant component could be indeed associated to them.}
Having established that our component is not related to the solar cycle and that 
it must be very close to being constant, we are left with a limited range of options. Particles of heliospheric origin have much softer spectra than CR, which means they cannot penetrate the shielding around the pn detector, with the possible exception of electrons, which have a much smaller cross section than protons and alpha particles. However, even if we were to accept the unlikely possibility  that supra-thermal electrons were responsible for the constant component, we would be faced with the challenge of explaining how a highly variable electron flux, variability of orders of several magnitudes is documented in the literature, might yield a constant component.

Another possibility is that the constant component is produced through activation. Activation is the process in which Cosmic Rays, mostly neutrons, excite nuclei, which later return to their ground state by emitting gamma rays. Since radioactive nuclei can exhibit half-lives ranging from small fractions of a second to several weeks or even months, this process allows for emission of gamma rays a long time after the material has been activated. After a few weeks or at most months in orbit, activation should produce a roughly continuous contribution to background. There are two reasons suggesting that activation is not responsible for the bulk of the constant component: 1) the half life for the vast majority  of the  excited nuclei is much shorter than the solar cycle, this implies that, even if with some delay, particle contribution due to activation should follow the solar cycle; 2) the contribution is expected to be small, in a recent study of the WFI detector to be flown on ATHENA, which for the purpose at hand is very similar to the EPIC pn, it has been shown that this process does not account for more than a few percent of the total background.

\begin{figure}
\centering
\includegraphics[width=9cm]{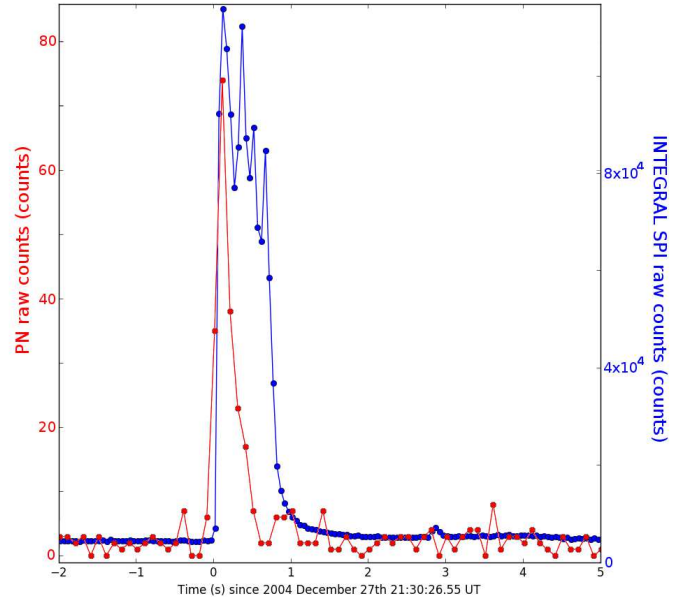}
\caption{Light curve of the giant flare from SGR 1806--20 as seen by {\it INTEGRAL/SPI} (18 keV--8 MeV, time bin size 0.05 s, the peak is saturated; \citealt{mereghetti05}) and {\it XMM-Newton}/pn (0.2--14 keV, time bin size 0.1 s). Time zero has been chosen following \citet{mereghetti05}. Contemporaneous flares have been registered also by MOS1 and MOS2 camera, although their time resolution is worse (2.6 s). Flare events are registered throughout the entire pn detector (both inFOV and outFOV) and its spectrum is consistent with a power law with $\Gamma\sim0.2$.
}
        \label{fig_intpn}
\end{figure}

If not particles, perhaps photons could be the primaries we are seeking. We have direct evidence that photons in the range 50--200 keV can penetrate the shielding around the pn detector. On 2004 December 27, a giant flare from the soft gamma repeater SGR 1806--20 was recorded by several observatories \citep[e.g.][]{hurley05,mereghetti05}. At precisely the same time, the EPIC pn and MOS detectors registered  an increase in the background rate of more than an order of magnitude (Fig.\,\ref{fig_intpn}).
Since the {\it XMM-Newton} telescopes where pointing at roughly 90 degrees from SGR 1806-20, we take this event as evidence  that hard X-ray photons penetrated all the way to the detector.
On a high Earth orbit, such as \emph{XMM-Newton}'s, the dominant source of hard X-ray photons is the integrated emission from celestial X-ray sources, the Cosmic X-ray Background (hereafter CXB). This component satisfies one of our main requirements, namely that it be constant. Indeed, as the sum of all sources in the sky, it does not vary in any appreciable way.  A related but independent property implied by the small scatter in the pn vs EPHIN relation, and satisfied by the CXB, is that this component must be essentially independent on the pn orientation in space. Indeed, a source coming from a specific region of the sky would produce a background rate on the pn that would depend on how it interacts with the asymmetric distribution of masses around the detector. Another important requirement is that the flux provided by the CXB be consistent with the intensity we measure.\\
This is what is found both from back of the envelope calculations, based on density and thickness of the shielding surrounding the EPIC detectors, and by detailed simulations on the ATHENA WFI detector \citep{vonkienlin18}, which bears strong similarities with the EPIC pn both in terms of shielding and detector. More specifically, \citet{vonkienlin18} have shown that, depending on the details of the shielding applied around the WFI detector, the CXB component can vary between $\sim$$0.6\times 10^{-3}$ and $\sim$$1.8\times 10^{-3}$ $\rm cts \, cm^{-2}\, s^{-1} \, keV^{-1}$. This is comparable  to the value we can deduce for our constant component. We point out that a more accurate estimate of the CXB contribution to the EPIC pn unconcentrated background is beyond the scope of this paper and may actually be hard to achieve given the limited documentation available for the EPIC cameras.

Our findings have profound implications for future X-ray experiments, particularly for the ATHENA WFI, which bears strong similarities with the EPIC pn. The tight correlation between EPHIN flux and pn background rate, as well as the detection of a secondary source of background events,  most likely associated to cosmic X-ray photons, provide  the first, important validations of simulation studies performed by the ATHENA WFI Background Working Group  \citep{vonkienlin18}. Moreover, the correlation between EPHIN and pn strongly suggests that a high energy particle monitor can contribute decisively to the reproducibility of the background for both experiments on ATHENA, i.e. WFI and XIFU.

\section{Summary} \label{summary}

We have analyzed the largest data set ever used for studies of the {\it XMM-Newton} EPIC/pn unconcentrated background. From the outFOV region of the detector, we collected $2.5\times10^6$ events in the 10-14 keV energy band using a total exposure time of 53 Ms on 1353 revolutions, spanning over twelve years (2000-2012).
Our analysis produced the following results:
\begin{itemize}
\item The region known as outFOV is in fact largely contaminated by concentrated photons and soft protons. Focusing on the soft proton contamination, its surface brightness radial profile follows a quasi-Lorentzian distribution whose normalization is proportional to the intensity of inFOV soft protons, but is constant in shape. The contamination from photons is consistent with having the same characteristics.
\item  A carefully selection of energy band and time periods allowed us to measure the unconcentrated background in the pn outFOV. We: 1) reduced the outFOV contamination of our total data set to less than 1\% of $sb_{sample}$ (the surface brightness evaluated over our entire sample) and 2) limited the contamination of all of the considered revolutions but a few outliers to less than 2.3\% (of $sb_{sample}$).
\item The unconcentrated background in the pn outFOV does not show significant spatial variations higher than $\sim$2\%, nor any dependence with the sub-mode or optical filters used; its time behaviour is anti-correlated with the solar cycle. Its surface brightness varies from $\sim3\times10^{-3}$ to $\sim7\times10^{-3}$ counts s$^{-1}$ cm$^{-2}$ keV$^{-1}$ over the considered period; the surface brightness evaluated over the entire period is $sb_{sample}=(5.118\pm0.003)\times10^{-3}$ counts s$^{-1}$ cm$^{-2}$ keV$^{-1}$.
\item We found a tight linear correlation between the pn and MOS2 outFOV surface brightness with an intrinsic scatter of 2.3\% (of $sb_{sample}$). This relationship permits the correct evaluation of the pn unconcentrated background of each exposure on the basis of MOS2 data, avoiding the use of the contaminated out-field of view region of pn.
\item We found a tight linear correlation between pn unconcentrated background and proton flux in the 630--970 MeV energy band, as measured by EPHIN instrument. The intrinsic scatter of this correlation is $<$3.5\% (of the median of pn values). This confirms that cosmic rays protons are one of the main contributors to the pn unconcentrated background accounting for 30\% to 70\% of it, depending on the time of observation during the solar cycle.
\item The non zero constant term in the EPHIN-pn relation implies that there is another source contributing to pn unconcentrated background. This agent does not depends on the solar cycle, it does not vary with time and it is roughly isotropic. After having ruled out several candidates, we found that hard X-ray photons of the CXB satisfy all known properties of this constant component. 
\item Our findings have profound implications for ATHENA. The tight correlation between EPHIN flux and pn background rate, as well as the detection of a secondary source of background events provide  the first, important validations of simulation studies performed on ATHENA. Moreover, the correlation between EPHIN and pn  strongly suggests that a high energy particle monitor can contribute decisively to the reproducibility of the background of both experiments on ATHENA.
\end{itemize}

\acknowledgements
We thank the scientific editor, the referee and their collaborators for the useful comments that really improved the paper in many ways.\\
This research has made use of data produced by the EXTraS project, funded by the European Union’s Seventh Framework Programme under grant agreement n. 607452. The scientific results reported in this article are based on observations obtained with XMM-Newton, an ESA science mission with instruments and contributions directly funded by ESA Member States and NASA.\\
The SOHO/EPHIN project is supported under grant 50 OC 1702 by the  German Bundesministerium f\"ur Wirtschaft through the Deutsches Zentrum f\"ur Luft- und Raumfahrt (DLR).\\
FG, SM, AdL and AT acknowledge support from INAF mainstream project ‘Characterizing the background of present and future X-ray missions’ 1.05.01.86.13.

\vspace{5mm}
\facilities{XMM, SOHO}

\software{astropy \citep{ast13},
  numpy \citep{oli06},
  matplotlib \citep{hun07},
  ltsfit \citep{cap13},
  scipy \citep{sci20},
  SAS \citep[v 14.0][]{gab04}}

\appendix

\section{Extraction of off-flare periods} \label{off_script}

We define the off-flare periods as the time intervals during which soft protons do not significantly contaminate the pn detector. Therefore, they comprise time in which only the constant component of the background is present.
To produce the GTIs of the off-flare periods, we rely on the inFOV light curves in the EXTraS database, which benefit from a better statistics than the light curve that we could obtain in the small outFOV region.\\
The EXTraS project \citep{del16} carefully evaluated and subtracted, through a complex modelling, point-like sources to obtain 500-s time bin background light
curves in the 0.2--12 keV energy band\footnote{https://www88.lamp.le.ac.uk/extras/archive}. Also, thanks to a better treatment of instrumental effects
-- such as the evaluation of CCD-dependent GTIs and the refinement of extraction region --
EXTraS light curves are better suited for background analysis than the standard results from SAS analysis.

Given the EXTraS inFOV background light curve as input, we implemented an algorithm to extract the off-flare periods. Before performing this analysis, we conservatively excluded all bins with a fractional exposure (the fractional CCD-weighted time of good exposure in the time bin) lower than 0.1 (thus an exposure lower than 50 s per bin) in order to have at least 30 counts per bin\footnote{Since the inFOV count rate changes with time, we extracted the number of registered events as a function of the fractional exposure for each bin. We verified that in each bin with fractional exposure $>$0.1 we registered at least 30 counts (while this is not true if we use a lower limiting fractional exposure).}.\\

\begin{figure*}[hbt!]
\centering
\includegraphics[width=0.75\textwidth]{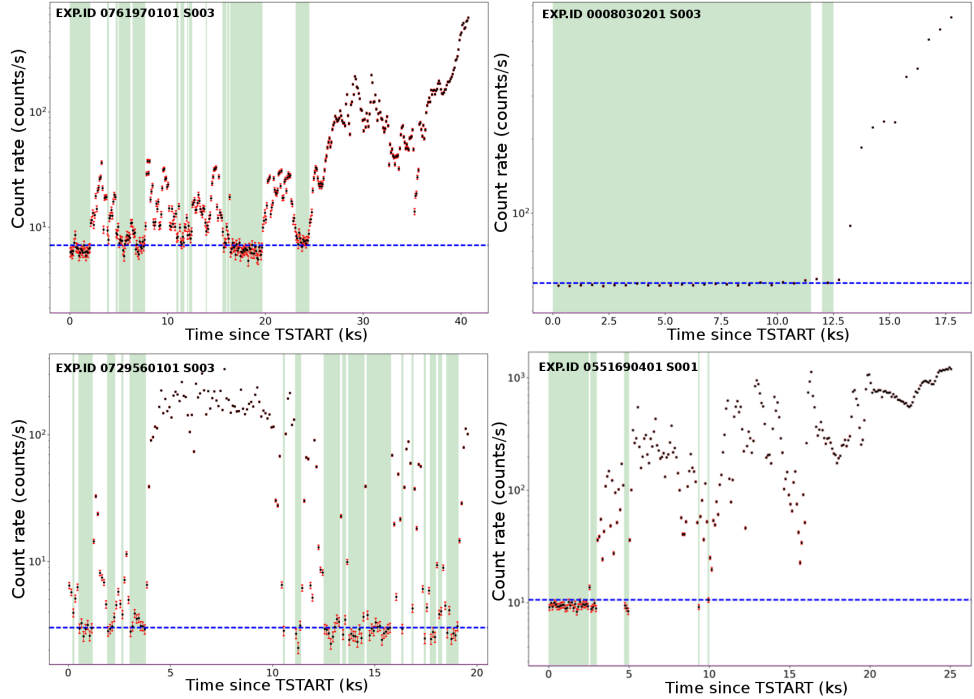}
\caption{EXTraS background light curve (from inFOV, 0.2--12 keV) for a sample of pn exposures. The original curve, named as reported in each panel, in fits format can be found within
  the EXTraS archive. Green areas show the resulting off-flare time period for each exposure, as obtained by our script and using $N=10$. $cr_{\mathrm{off}}$ is shown in dashed-blue.}
\label{fig_offtime}
\end{figure*}

The aim of our algorithm is to maximize the duration of the off-flare period, where the background light curve is consistent with a constant. This is equivalent to finding the highest possible constant value $cr_{\mathrm{off}}$ that fits the uncontaminated part of the light curve at a given confidence level. More specifically, we iteratively set a value of $cr_{\mathrm{off}}$ and exclude the time bins whose count rate $cr_{\mathrm{i}}$ shows the highest positive deviation $\delta_{\mathrm{i}}=(cr_{\mathrm{i}}-cr_{\mathrm{off}})/\sigma_{\mathrm{i}}$ ($\sigma_{\mathrm{i}}$ being the statistical error on the count rate of the $i-$th time bin). We iterate this procedure with lower and lower values of $cr_{\mathrm{off}}$, until all remaining bins are consistent with a constant at a given confidence level, parametrized by the input value $N_\sigma$ (see below). These time bins form the off-flare period. A few examples of light curves filtered with our algorithm are shown in Fig. \ref{fig_offtime}.

\begin{figure}
\centering
\includegraphics[width=9cm]{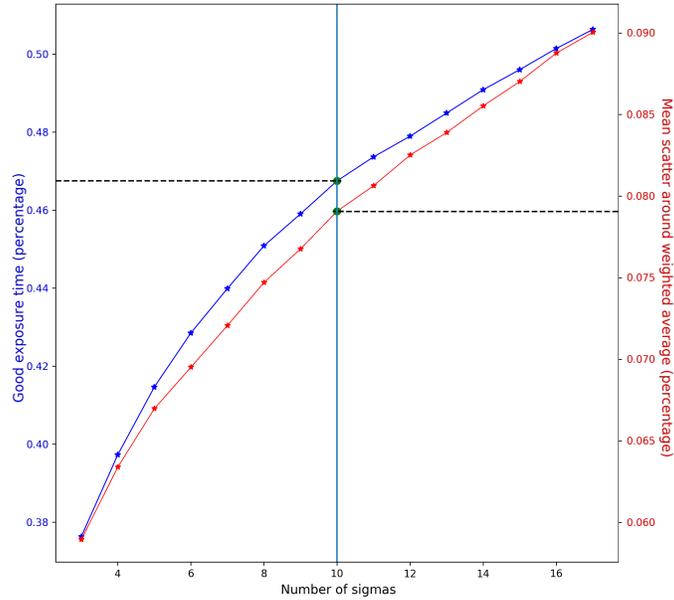}
\caption{In blue, percentage of exposure time as a function of the parameter $N_{\sigma}$ in our script. In red, the mean of the spread of each exposure (normalized for their weighted averages) for different values of N$\sigma$. All the parameters are evaluated over our Main sample.
}
        \label{fig_sigmas}
\end{figure}

To obtain the best selection of off-flare periods, we have to choose wisely the confidence level used to check the consistency of the count rates with the constant, which is expressed by the input $N_{\sigma}$.
To do that we ran our analysis using different values of $N_{\sigma}$, from
3 to 17.  As an indicator of residual soft proton contamination in each exposure after filtering with a given $N_\sigma$, we use the scatter of the count rates around their weighted average. We then compare the mean of the scatters over all observations as a function of $N_{\sigma}$ in  Figure \ref{fig_sigmas}.
We chose $N_{\sigma}$ = 10 as a reasonable compromise between contamination minimization and exposure time maximization.

\begin{figure}
\centering
\includegraphics[width=9cm]{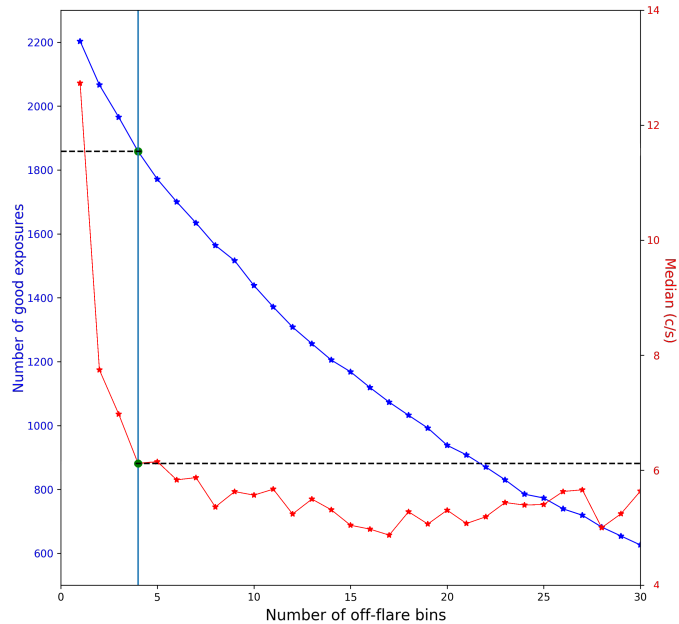}
\caption{In blue, the percentage of off-flare exposure time as a function of the minimum off-flare bins per exposure.
In red, the median of count rates of the distributions of exposures with N off-flare time bins as a function of N. All the parameters are evaluated over our Main sample.
}
        \label{fig_minofftime}
\end{figure}

Using our script, by construction every curve has at least one off-flare time bin, even for exposures always contaminated by SP: the time bin with the lowest count rate in the light curve. Thus, we have to set a minimum number of good bins: below this value, the entire exposure is conservatively rejected.
As for the choice of $N_{\sigma}$, the minimum number of bins of off-flare time is chosen as a compromise between contamination minimization and exposure time maximization.
To estimate the contamination in the exposures as a function of the number of good bins, we divide them in subsamples based on the number of good bins.  We then compute the median  of the distributions of off-flare count rates for each sub-sample\footnote{Since contaminated exposures can have a count rate that is orders of magnitude higher than uncontaminated ones,  single exposures can dominate the mean of the entire distribution. Therefore, we chose to use the median.}.
In Figure \ref{fig_minofftime} we show these values as a function of the number of
off-flare time bins per exposure. Since the contamination decreases quickly with increasing number of off-flare bins and reaches a plateau
around 4, we chose to reject all exposures with less than 4 off-flare time bins.
Moreover, to exclude low-statistic revolutions in our Main data set, we shall only consider revolutions with at least 10 ks of off-flare time.

\section{Modelization of OutFOV contamination from soft protons} \label{app:cont}

\begin{figure*}
  \includegraphics[width=0.75\textwidth]{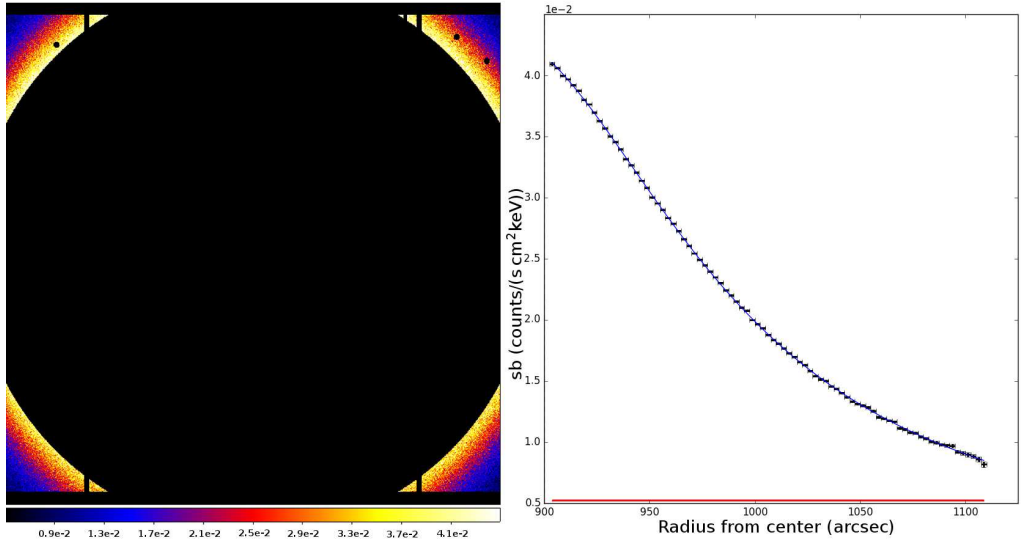}
  \caption{{\it Left Panel:} Surface brightness map defined as in Section \ref{imaging}, for our on-flare sample, in detector coordinates and units of counts s$^{-1}$ cm$^{-2}$ keV$^{-1}$.
  {\it Right Panel:} Surface brightness profile from the geometrical center of pn detector with a 5'' step, computed following Equation 6, computed for our on-flare sample ($\sim10^7$ counts). We used a step of 2.5''. Effects related to OoT events, area and time rescale are
  considered as described in Section \ref{imaging}. In red we report the predicted off-flare brightness. In blue we report the best fit described by Equation B.1.
}
\label{fig_histoflares}
\end{figure*}

Since SP are the main source of contamination in the outFOV region for most of the {\it XMM-Newton} exposures, we want to study the radial dependence of the contamination and evaluate its impact in our main data set (Section \ref{imaging}).
To do this we decided to study the complementary contaminated data set, because it has more statistics than our clean data set, describe the radial dependence in the outFOV region with an empyric model that we then rescale to estimate the residual contamination in the clean data.\\
We made the same selection as in Section \ref{data}, but we kept only the complementary on-flare times.
We chose to ignore periods with inFOV rates higher than 500 counts s$^{-1}$ in order to minimize pile-up effects.\\
Using the same event selection as in Section \ref{evt_sel}, we obtain a data set comprising $1.06\times10^7$ counts in the outFOV region
($9\times10^5$ are expected to be OoT counts) over a 40.0 Ms exposure time.
Using this huge data set, we built the radial surface brightness profile of the outFOV region during on-flare periods,
following the same method reported in Section \ref{imaging}. In Figure \ref{fig_histoflares}, we report the surface brightness map and the surface brightness radial profile.
The observed surface brightness during on-flare periods is maximum near the edge of the FoV and decreases steadily with increasing radius; the entire outFOV area is highly contaminated by soft protons, as seen in Section \ref{cont_pre} for small sub-samples.
We assume that the unconcentrated background has a flat brightness profile, while the profile variation is only due to the contamination.
Both for the MOS2 camera \citep{mar17} and for pn camera (Section \ref{imaging}), this is a fair assumption.
We evaluated the surface brightness due to unconcentrated background during on-flare periods of our entire sample to be $sb_{\mathrm{onflare}}=5.21\times10^{-3}$ counts s$^{-1}$ cm$^{-2}$ s$^{-1}$ using the surface brightness during off-flare periods of our entire sample ($sb_{sample}$ from Section \ref{lc_creation}).
We empirically found that the brightness profile in Figure \ref{fig_histoflares} follows a Lorentzian model
plus the constant $b_{\mathrm{onflare}}$, with a corrective factor depending only on the Lorentzian normalization:\\

\begin{equation}
b(r) = N \frac{w^2}{(r-r_0)^2+w^2} - 0.1158 N + b_{\mathrm{onflare}}
\end{equation} \label{eq_brig_flares}

\noindent where the width is $w = (113.4\pm0.3)''$ and the maximum is located at $r_0 = (875.6\pm0.6)''$ (from the geometrical center of the pn detector).
Despite the huge amount of events ($\sim10^7$), the fit is fully acceptable, with a null hypothesis probability (n.h.p.) of 0.13.

We checked the shape and/or parameters of the radial brightness profile for a dependance on a) intensity of SPs (counts s$^{-1}$), b) time or c) angle.\\
a) We divided our sample for different levels of flares to obtain ten data sets with a similar statistics ($\sim10^6$ counts).
We built the brigthness profiles of the outFOV region for each of them.
Equation B.1 always provides a good fit (n.h.p. $>0.1$), where width $w$ and maximum $r_0$ are always compatible within 2$\sigma$ with
the ones obtained for the entire sample. The only variable parameter is the normalization $N$, that increases proportionally to the inFOV flare rate.\\
b) We divided our sample in twelve sub-sample, one for each year of observation. Again, Equation B.1 fits well all the profiles and we have no evidence
of variation of $w$ and $r_0$.\\
c) We extracted and fitted separately the radial brightness profiles of the four quadrants (corners) of pn.
Although the fit is still good for each of them (n.h.p. $>0.02$), we find significant differences in
the four best-fitted sets of parameters ($>5\sigma$). The obtained radial brightness profiles and best
fits are reported in Figure 14 and Table 2. The normalization $N$ is maximum in the quadrant nearest to the boresight position and then
decreases with the distance from it.
This seems to point to a dependance of the contamination to the boresight position and/or the proton vignetting.

We conclude that in the 10--14 keV band the surface brightness profile of the contamination due to soft protons in the pn outFOV region follows the empirical
Equation B.1, where only the normalization $N$ varies, increasing with the inFOV flare rate. The contamination is slightly angular-dependant
but it always follows the empirical distribution of Equation B.1.\\
We note that the radial brightness profile of exposures highly contaminated by photons (``photon sample'' in Section \ref{cont_pre}),
is compatible with the same equation, although we rely on a much lower statistics ($\sim10^4$ counts).

\begin{figure*}
\begin{floatrow}
\ffigbox{%
  \includegraphics[width=0.4\textwidth]{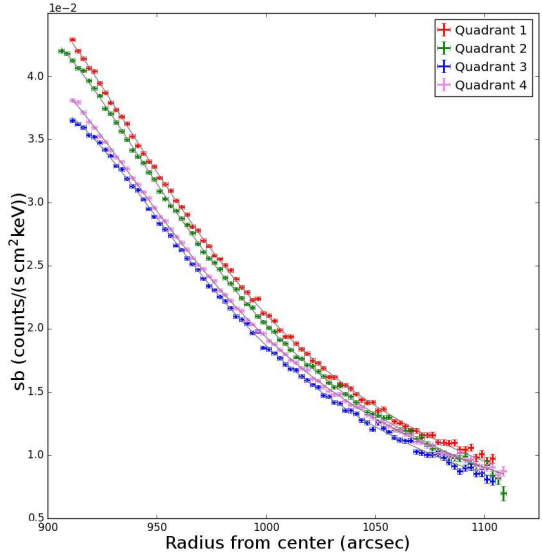}
}{%
  \caption{Surface brightness profiles of outFOV region during on-flare periods over our entire data set, divided by pn quadrant. We used a step of 2.5''.
  All the effects related to OoT events, area and time rescale are
  considered as described in Section \ref{imaging}. In grey we report the best fits described by Equation B.1.}%
}

\capbtabbox{%
  \begin{tabular}{lcccc} \hline
  Quad. & n.h.p & $N$ & $w$ & $r_0$ \\
    & & 10$^{-2}$ c s$^{-1}$cm$^{-2}$keV$^{-1}$ & " & " \\ \hline
  1 & 0.026 & 4.90$\pm$0.05 & 117.4$\pm$0.7 & 868$\pm$2\\
  2 & 0.211 & 4.55$\pm$0.03 & 112.2$\pm$0.5 & 875$\pm$1\\
  3 & 0.049 & 3.80$\pm$0.03 & 108.0$\pm$0.6 & 885$\pm$1\\
  4 & 0.989 & 4.23$\pm$0.03 & 116.3$\pm$0.5 & 871$\pm$1\\ \hline
  \end{tabular}
  \vspace{2.8cm}
}{%
  \caption{Parameters of the best fits of the brightness profiles of outFOV region during on-flare periods over our entire data set, divided by pn quadrant.
  The model is reported in Equation B.1.}%
}
\end{floatrow}
\end{figure*}

\section{SAS event selection} \label{app:SAS}

We report here the event selection from Section \ref{evt_sel} in SAS notation.

We define as outFOV area:\\
{\tt ((FLAG \& 0x10000)!=0) \&\& !((DETX,DETY) IN circle(-2203.00,-1107.00,18101.00)) \&\& !(DETY <= -16587) \&\& !(DETY in [-1227:-987]) \&\& !(DETY >= 14373) \&\& !(DETX <= -18243) \&\& !(DETX in [-13143:-12843]) \&\& !(DETX in [-7763:-7443]) \&\& !(DETX in [-2363:-2043]) \&\& !(DETX in [3037:3337]) \&\& !(DETX in [8417:8737])\&\&!(DETX >= 13817)}\\
We make use of the following pattern and flag selection:\\
{\tt (PATTERN<=4) \&\& (FLAG \& 0x1)==0 \&\& (FLAG \& 0x20)==0 \&\& (FLAG \& 0x40)==0 \&\& (FLAG \& 0x100)==0 \&\& (FLAG \& 0x20000)==0 \&\& (FLAG \& 0x80000)==0 \&\& (FLAG \& 0x100000)==0 \&\& (FLAG \& 0x200000)==0 \&\& (FLAG \& 0x400000)==0 \&\& (FLAG \& 0x800000)==0}\\
We make use of the following energy selection:\\
{\tt (PI in [10000:14000])}\\
We excluded the following bright pixels and columns:\\
{\tt !((DETX,DETY) IN circle(12957.0,11375.5,201.0) || (DETX,DETY) IN circle(11004.5,12943.0,201.0) || (DETX,DETY) IN circle(-14953.0,12423.0,201.0) || (DETX,DETY) IN box(7677,6753,121,7781,0))}

\bibliography{biblio}{}
\bibliographystyle{aasjournal}

\end{document}